\documentclass[floatfix,aps,unsortedaddress,superscriptaddress,showpacs]{revtex4}

\usepackage{graphicx}
\usepackage{amsmath}
\usepackage{amssymb}
\usepackage{bm}
\newcommand*{\brho}{{\bm{\rho}}}

\newcommand*{\rk}{{\rm{\bf k}}}

\newcommand*{\be}{\begin{equation}}
\newcommand*{\ee}{\end{equation}}
\newcommand*{\bay}{\begin{eqnarray}}
\newcommand*{\eay}{\end{eqnarray}}
\newcommand*{\SE}{Schr\"odinger equation }

\begin{document}

\title{The leading asymptotic terms of the three-body Coulomb scattering wave function
}
\author{A. M. Mukhamedzhanov}
\affiliation{Cyclotron Institute, Texas A\&M University, College
Station, TX 77843, USA }

\author{A. S. Kadyrov}
\affiliation{Centre for Atomic, Molecular and Surface Physics, Division 
of Science and
Engineering, Murdoch University, Perth 6150, Australia}

\author{F. Pirlepesov}
\affiliation{Cyclotron Institute, Texas A\&M University, College 
Station, TX 77843, USA }

\date{\today}

\begin{abstract}
The asymptotic wave function derived by Alt and Mukha\-med\-zhanov 
[Phys. Rev. A 47, 2004\,(1993)] and Mukha\-med\-zhanov and 
Lieber
[Phys. Rev. A 54, 3078\,(1996)] has been 
refined in the region where the pair $
(\beta,\gamma)$ remains 
close to each other while the third particle $\alpha$ is far away from 
them $(\rho_{\alpha} \to \infty$, $r_{\alpha}/\rho_{\alpha}\to 0)$. The 
improved wave function satisfies the Schr\"odinger equation up to the 
terms of order $O(1/{\rho_{\alpha}}^{3})$, provides the leading asymptotic terms 
of the three-body scattering wave function with Coulomb interactions and gives further 
insight into the continuum behavior of the three-charged-particle wave function and 
helps to obtain $3\to 3$ scattered wave. This opens up further 
ways of solving and analysing the three-body Schr\"odinger equation by 
numerical means.
\end{abstract}
\maketitle

\section{Introduction}

The quantum
dynamics of three charged particles is 
described by Schr\"odinger's equation which should be supplemented by 
proper boundary conditions. 
Merkuriev and Faddeev \cite{FM93} claimed that the solution 
of this equation exists and is unique if the boundary conditions are 
known in all asymptotic regions. There are two types of the three-body 
scattering wave functions. The first type evolves from an initial three-body incident 
wave describing three incident particles in continuum. 
The second type of the three-body scattering wave function evolves from a two-body 
incident wave corresponding to collision of a two-body bound state with a third particle.

The three-body incident wave represents the leading asymptotic 
terms of the total three-body scattering wave function \cite{FM93,
AM93,ML96}. The knowledge of the three-body incident wave is important 
for many reasons: as a leading term of the three-body wave function
it can be used in calculations of the breakup matrix elements if the kinematics
is such that the asymptotic region gives the leading contribution; 
the knowledge of the three-body incident wave is necessary in direct 
solution of the three-body Schr\"odinger equation for the 
scattering wave function of the first type. 
The asymptotic behavior of the three-body incident wave depends on the asymptotic 
region under consideration.
In the asymptotic region where two particles, for example  $\beta$ and $\gamma$, are 
close to each other and far away
from the third particle $\alpha$, the three-body incident wave can be written as 
an asymptotic series in powers $1/\rho_{\alpha}$, where $\rho_{\alpha}$ is the 
distance between the c.m. of the system $(\beta , \gamma)$ and the third 
particle $\alpha$. The leading asymptotic terms $O(1)$ and $O(1/\rho_{\alpha})$ 
of the three-body incident wave for 
charged particles have been obtained 
analytically in \cite{AM93,ML96}. These asymptotic terms 
satisfy the Schr\"odinger equation up to $O(1/\rho_{\alpha}^{2})$. 
In this work we will derive all the leading asymptotic terms of the three 
charged particles incident wave of order $O(1/\rho_{\alpha}^{2})$. 
Combined with the previously derived terms 
of order
$O(1)$ and $O(1/\rho_{\alpha})$ \cite{AM93,ML96}, 
they provide the asymptotic solution of the Schr\"odinger equation for three charged particles in continuum up to terms 
$O(1/\rho_{\alpha}^{3})$. The terms $O(1/\rho_{\alpha}^{2})$ satisfy first order 
differential equations.
It is worth mentioning that the terms $O(1/\rho_{\alpha}^{3})$ are the next 
order terms compared to the three-body scattered wave which is $O(1/R^{5/2})$, 
where $R$ is the hyperradius. This 
term, as well as the leading-order asymptotic terms of the three-body 
wave function of the second type have been given in Refs.~\cite{KMS03,KMSBP03}. 
Practical ways of extracting the scattering and breakup amplitudes using these asymptotic
wave functions have been presented in Refs.~\cite{KMSB03l,KMSB04}.

The paper is organized in the following way. In Section I we introduce the three-body 
nomenclature and give the statement of the problem. In Section II we recall 
some of the important relations relevant to two-body scattering. In 
Sections III-V we present asymptotic solutions of the three-body 
Schr\"odinger equation in all orders which can be obtained analytically with the 
asymptotic method. Finally, Section F concludes the paper.

\section{Statement of the problem}

We consider a non-relativistic three-body problem for charged 
particles of mass $m_{\alpha}$ and charge $z_{\alpha}$, $\alpha=1,2,3$ 
in the continuum state. We follow the notations used in Ref. \cite{AM93}.
The Greek letters stand for constituent
particles of the three-body system or for the pair of two other particles. For example, $\alpha$ labels the particle or the pair $\beta + \gamma$. 
Such a supplemental notation is customary in few-body physics. 
The following conventional notations for the two body quantities are also
used: $A_{\alpha}\equiv A_{\beta\gamma}$, where $\alpha\neq \beta\neq 
\gamma$. 
The Jacobi coordinates are determined as follows: 
${\mathbf{r}}_{\alpha}$ is the relative coordinate between 
particles $\beta$ and $\gamma$, and ${\mathbf{k}}_{\alpha}$ is its  
canonically conjugated momentum.  
$\mu_{\alpha}=\frac{m_{\beta}m_{\gamma }}{m_{\beta \gamma}}$ is 
their reduced mass, $m_{\beta \gamma}= m_{\beta} + m_{\gamma}$. 
Similarly, ${\brho}_{\alpha }$ is the relative 
coordinate between the c.m. of the pair $(\beta,\gamma)$ and 
particle $\alpha$, and 
${\mathbf{q}}_{\alpha }$ is its canonically conjugated relative 
momentum. 
$M_{\alpha}= m_{\alpha}\,m_{\beta\gamma}/M$,
$M=\sum\limits_{\nu=1}^{3}m_{\nu}$ is total mass of the three-body 
system. 
There are 
three sets of Jacobi coordinates 
${\mathbf{r}}_{\nu},{\brho}_{\nu }$, where 
$\nu=\alpha,\beta,\,\gamma$.
We frequently need the relations between the coordinates, and
conjugate momenta for a channel $\nu=\beta,\gamma$ and
the corresponding $\alpha-$channel variables. They are given by the 
following relations
\begin{eqnarray}
\left( \begin{array}{c}{\brho}_{\nu} \\ 
{\mathbf{r}}_{\nu}\end{array}\right) =\left(\begin{array}{cc}
-\frac{m_{\alpha}}{M-m_{\nu}} & \epsilon_{\nu\alpha}\frac{\mu_{\nu}}{%
M_{\alpha}} \\ 
-\epsilon_{\nu\alpha} & -\frac{m_{\nu}}{m_{\beta\gamma}}
\end{array}
\right) \left( 
\begin{array}{c}
{\brho}_{\alpha} \\ 
{\mathbf{r}}_{\alpha}
\end{array}
\right) 
\end{eqnarray}
\begin{eqnarray}
\left( 
\begin{array}{c}
{\mathbf{q}}_{\nu} \\ 
{\mathbf{k}}_{\nu}
\end{array}
\right) =\left( 
\begin{array}{cc}
-\frac{m_{\nu}}{m_{\beta\gamma}} & \epsilon_{\nu\alpha} \\ 
-\epsilon_{\nu\alpha}\frac{\mu_{\alpha}}{M_{\nu}} & -\frac{m_{\alpha}}{%
M-m_{\nu}}
\end{array}
\right) \left( 
\begin{array}{c}
{\mathbf{q}}_{\alpha} \\ 
{\mathbf{k}}_{\alpha}
\end{array}
\right), 
\end{eqnarray}
where $\nu=\beta,\gamma$ and the antisymmetric symbol 
$\epsilon_{\alpha\nu}=-\epsilon_{\nu\alpha}$, with 
$\epsilon_{\alpha\nu}=1$ for 
$(\alpha,\nu)$ being a cyclic permutation of $(1,2,3)$, and 
$\epsilon_{\alpha\alpha}=0$.
The motion of the three particles is described by the \SE in the 
configuration space
\begin{eqnarray}
\{E-T_{{\mathbf{r}}_{\alpha}}-T_{{\brho}_{\alpha }}
-V\}\Psi_{{\mathbf{k}}_{\alpha}{\mathbf{q}}_{\alpha}}^{(+)}
({\mathbf{r}}_{\alpha},{\brho}_{\alpha})=0,
\label{ThSE}
\end{eqnarray}
where $V=\sum\limits_{\nu=1}^{3}V_{\nu}$, 
$V_{\nu}=V_{\nu}^{C}({\mathbf{r}}_{\nu})+V_{\nu}^{N}({\mathbf{r}}_{\nu}
)$.
The Coulomb potential is given by
$V_{\alpha}^{C}({\mathbf{r}}_{\alpha})=\frac{Z_{\beta}\,Z_{\gamma}\,e^{2}}
{r_{\alpha}}$, $Z_{\nu}\,e$ is the charge of particle $\nu$. 
Similarly, $V_{\nu}^{N}$ is the nuclear potential between the particles 
of the $\nu-$ pair, where $\nu=\alpha,\beta,\gamma$.
$T_{{\mathbf{r}}_{\alpha}}
=-\frac{\bigtriangleup _{{\mathbf{r}}_{\alpha}}}{2\mu_{\alpha}}$, 
is the kinetic energy operator for the relative motion of particles 
$\beta$ and $\gamma$, and 
$T_{{\brho}_{\alpha}}=-\frac{\bigtriangleup_{{\brho}_{\alpha}}}
{2M_{\alpha}}$ is the kinetic energy operator for the relative motion of 
particle $\alpha$ and the center of mass of the pair $(\beta,\gamma)$, 
respectively.

Our aim is to derive the asymptotic behavior of three-body incident wave up to terms $O(1/\rho_{\alpha}^{3})$ in the asymptotic region $\Omega_{\alpha}$, 
where $r_{\alpha}/\rho_{\alpha} \to 0$ and $\rho_{\alpha} \to \infty$. 
This incident wave provides the leading asymptotic terms of the three-body scattering wave function of the first type.
General asymptotic behavior of the three-body scattering wave 
function is given by  \cite{FM93}
\begin{eqnarray}
\Psi_{{\rm{\bf{k}}}_{\alpha},{\rm{\bf{q}}}_{\alpha}}^{(+)}\approx  
{\tilde \Psi}_{{\rm{\bf{k}}}_{\alpha},{\rm{\bf{q}}}_{\alpha}}^{(+)}
+ \sum\limits_{\nu = \alpha,\,\beta,\,\gamma} \,
\varphi_{\nu}({\rm{\bf{r}}}_{\nu})\frac{{\cal{M}}^{(\nu)}_{3\rightarrow 
2}}{\rho_{\nu}}\,e^{i\,q_{\nu}\,{\rho}_{\nu}
-i\overline{\eta}_{\nu}\ln(2q_{\nu}\,\rho_{\nu})}   \nonumber\\
+\frac{{\cal{M}}_{3\rightarrow 3}}{R^{5/2}}\,
e^{i\kappa R-i\,\lambda_{0}\,\ln(2\kappa\,R) }.
\label{eq3binc1}
\end{eqnarray}
Here the first term is the incident three-body wave, the sum over $\nu$ 
provides the two-body outgoing scattered waves and corresponds to the $3 \to 2$
processes. The last term descibes the outgoing three-body scattered wave. 
Also, $\overline{\eta}_{\alpha}= (Z_{\beta} + Z_{\gamma})\,Z_{\alpha}\,e^{2}\,M_{\alpha}/q_{\alpha}$ is the Coulomb parameter for the Coulomb interaction between
particles $\alpha$ and the center of mass of the system $\beta + \gamma$;
the Coulomb pasrameter $\lambda_{0}$ is determined in Ref. \cite{KMSBP03}.  
We use the system of units such that $\hbar=c=1$.  
Formally we can determine the incident wave as an 
asymptotic difference 
\begin{eqnarray}
{\tilde \Psi}_{{\rm{\bf{k}}}_{\alpha},{\rm{\bf{q}}}_{\alpha}}^{(+)}
\approx  \Psi_{{\rm{\bf{k}}}_{\alpha},{\rm{\bf{q}}}_{\alpha}}^{(+)}
- \sum\limits_{\nu = \alpha,\,\beta,\,\gamma} \,
\varphi_{\nu}({\rm{\bf{r}}}_{\nu})\frac{{\cal{M}}^{(\nu)}_{3\rightarrow 
2}}{\rho_{\nu}}\,e^{i\,q_{\nu}\,{\rho}_{\nu}
-i\overline{\eta}_{\nu}\ln(2q_{\nu}\,\rho_{\nu})}   \nonumber\\
-\frac{{\cal{M}}_{3\rightarrow 3}}{R^{5/2}}\,
e^{i\kappa R-i\lambda_{0}\,\ln(2\kappa\,R)}.
\label{eq3binc10}
\end{eqnarray}
From this equation it is clear that the three-body incident wave is a
part of the full wave function, which does not contain the outgoing two- 
and three-body scattered waves. For better understanding of the three-body 
incident wave we consider first the two-body case.

\section{Asymptotic two-body scattering wave function}

We will be referring to the two-body Coulomb scattering throughout this 
work. Therefore we present here some important relations for the two-body 
scattering. 
Let us consider two charged particles with mass $m_{i}$ and  
charge $Z_{i}\,e,\;i=1,2$,  interacting via the pure Coulomb potential 
$V=Z_{1}\,Z_{2}\,e^{2}/r$.  
Scattering of two particles is described by the \SE 
\begin{eqnarray}
\{E-H\}\psi_{{\mathbf{k}}}^{(+)}({\mathbf{r}})=0,
\label{TbSE}
\end{eqnarray}
where ${\eta}=Z_{1}\,Z_{2}\,e^{2}\,\mu/k$ is the Coulomb parameter, 
$E=k^{2}/(2\,\mu)$ is the relative kinetic energy of the 
interacting particles $1$ and $2$,
$H=-\bigtriangleup_{{\mathbf{r}}}/(2\,\mu)+V$ is two body 
Hamiltonian,
and $\mu=m_{1}\,m_{2}/(m_{1}+m_{2})$ is reduced mass of particles 
$1$ and $2$.  
For the pure Coulomb interaction case Eq.(\ref{TbSE}) can be solved 
analytically. Substituting  
\begin{eqnarray}
\psi_{{\mathbf{k}}}^{(+)}({\mathbf{r}})
=e^{i{\mathbf{k}}\cdot {\mathbf{r}}}\, N\;{}_{1}F_{1}(-i\eta ,1,i\zeta ),       \label{twbclscwf1}
\end{eqnarray}
into Eq.(\ref{TbSE}) gives the differential equation for the confluent 
hypergeometric function (also called the Kummer function)
\begin{eqnarray}
\lbrack \frac{\bigtriangleup _{{\mathbf{r}}}}{%
2\mu }+\frac{i{\mathbf{k}}\cdot \bigtriangledown _{{\mathbf{r}}}}{\mu 
}-V]\,{}_{1}F_{1}(-i\eta ,1,i\zeta )=0,  \label{hyprgfnct1}
\end{eqnarray}
$N=e^{-\pi{\eta }/2}\Gamma(1+i{\eta})$ is the normalization 
factor, and parabolic coordinate $\zeta=kr-{\mathbf{k}}\cdot{\mathbf{r}}$. 
${}_{1}F_{1}(-i\eta ,1,i\zeta )$ is called the Kummer function because Eq. (\ref{hyprgfnct1}) rewritten in terms 
of $z=i\,\zeta$ becomes the Kummer differential equation \cite{LL85,KF63}:
\begin{equation}
z\,\frac{{\rm d}^{2}\,_{1}F_{1}(a,c;z)}{{\rm d}z^{2}}  +  (c -z)\,\frac{{\rm d}\,_{1}F_{1}(a,c;z)}{{\rm d}z}  - a\;_{1}F_{1}(a,c;z)=0. \label{kummereq1}
\end{equation}
Here $a=-i\eta,\,c=1$. Note that the Kummer function $_{1}F_{1}(a,c;z)$ is a regular solution at $\zeta=0$ (or $r=0$) of the Kummer equation.
Correspondingly, $\psi_{{\mathbf{k}}}^{(+)}({\mathbf{r}})$ given by 
Eq. (\ref{twbclscwf1}) is the normalized regular solution of the two-body Coulomb 
scattering problem. The Kummer function can be 
expressed in terms of the Whittaker functions 
$W_{\lambda,\mu}(z)$ using Eqs. (9.220.3) and (9.233.2) \cite{GR80}:
\begin{eqnarray}
{}_{1}F_{1}(-i\eta ,\,1;i\,\zeta)= \frac{1}{\Gamma(1+ i\,\eta)}\,e^{\pi\,\eta}\,\frac{1}{(i\,\zeta)^{1/2}}\,e^{i\,\zeta/2}\,W_{1/2+i\,\eta}(i\,\zeta) \nonumber\\
+ \;\frac{1}{\Gamma(-i\,\eta)}\,e^{-i\,\pi(1/2+i\,\eta)}\,\frac{1}{(i\,\zeta)^{1/2}}\,e^{i\,\zeta/2}\,W_{-1/2-i\,\eta,0}(e^{-i\,\pi}\,i\,\zeta).
\label{hprgwhitfnc1}
\end{eqnarray}
Each term of Eq. (\ref{hprgwhitfnc1}) also satisfies the Kummer differential 
equation (\ref{kummereq1}) providing a singular solution. 
Substituting Eq. (\ref{hprgwhitfnc1}) into the Kummer equation leads to the 
Whittaker differential equation for each term:
\begin{equation}
\frac{ {\rm d}^{2}W_{\lambda,0}(z) }{ {\rm d}z^{2}}  +  (-\frac{1}{4} +
\frac{\lambda}{z} + \frac{\frac{1}{4} - \nu^{2}}{z^{2}})\,W_{\lambda,0}(z)=0. 
\label{whittaker1}
\end{equation}
Here $\lambda= \pm(1/2+ i\,\eta)$ and $z=\pm i\,\zeta$. 
Evidently that both Whittaker functions in Eq. (\ref{hprgwhitfnc1})
satisfy the same Whittaker equation because it is invariant under simultaneous transformation $z \to -z,\,\lambda \to -\lambda$. 
Coming back to the normalized regular solution of the Schr\"odinger equation 
we can present it as a sum of two singular solutions:
\begin{eqnarray}
\psi_{{\mathbf{k}}}^{(+)}({\mathbf{r}})
=e^{i{\mathbf{k}}\cdot {\mathbf{r}}}\, N\,{}_{1}F_{1}(-i\eta ,1,i\zeta ) \nonumber\\
= \psi_{{\mathbf{k}}}^{(0)}({\mathbf{r}}) + 
\psi_{{\mathbf{k}}}^{(sc)}({\mathbf{r}}).  
\label{twbclscwf2}
\end{eqnarray}
The first singular solution, as we will see below, is the incident wave
\begin{equation}
\psi_{{\mathbf{k}}}^{(0)}({\mathbf{r}})= e^{i{\mathbf{k}}\cdot \mathbf{r}}\,{\cal{F}} ^{(1)}(\zeta ), 
\label{incwv1}
\end{equation}
and the second singular solution is the scattered wave
\begin{equation}
\psi_{{\mathbf{k}}}^{(sc)}({\mathbf{r}})= e^{i{\mathbf{k}}\cdot {\mathbf{r}}}\,{\cal{F}} ^{(2)}(\zeta ), 
\label{sctwv1}
\end{equation}
\begin{eqnarray}
{\cal{F}}^{(1)}(\zeta )=e^{\frac{\pi \eta }{2}}(i\zeta )^{-\frac{1%
}{2}}e^{i\frac{\zeta }{2}}W_{i\eta +\frac{1}{2},0}(i\zeta ),  
\label{whit1}
\end{eqnarray}
\begin{eqnarray}
{\cal{F}}^{(2)}(\zeta )=-i\frac{\Gamma (1+i\eta )%
}{\Gamma (-i\eta )}e^{\frac{\pi \eta }{2}}(i\zeta 
)^{-\frac{1}{2}}e^{i\frac{%
\zeta }{2}}W_{-i\eta -\frac{1}{2},0}(e^{-i\,\pi}\,i\zeta ).    \label{whit2}
\end{eqnarray}
Evidently that for $\eta=0$ the incident wave becomes the plane wave 
$e^{i{\mathbf{k}}\cdot \mathbf{r}}$ and the scattered wave just disappears. It follows from Eq. (9.227) \cite{GR80} that ($|z| >0$)
\begin{equation}
W_{1/2,0}(z) = e^{ - z/2} z^{1/2}\int\limits_0^\infty  {dte^{ - t} \frac{1}{{t + z}}}.  
\label{whitasexp1}
\end{equation}
and from Eq. (6.922.2) \cite{GR80}
\begin{equation}
W_{ - 1/2,0} (z) = e^{ - z/2} z^{1/2} \int\limits_0^\infty  {dte^{ - t} \frac{1}{{t + z}}}. 
\label{whitsc2}
\end{equation}
Taking into account the asymptotic behaviour  at $|z| \rightarrow \infty $ of the Whittaker function, Eq. (9.227) \cite{GR80}, 
\begin{eqnarray}
W_{\lambda ,0}(z ) \stackrel{|z| \to \infty}{=}z^{\lambda }e^{-z /2}\left[ 1-%
\frac{(\lambda -1/2)^{2}}{z }+O(\frac{1}{z^{2}})\right],    
\label{whitasym1} 
\end{eqnarray}
we derive the asymptotic behavior of ${\cal{F}}^{(1)}(\zeta 
)$:
\begin{eqnarray}
{\cal{F}}^{(1)}(i\zeta )\stackrel{|\zeta| \to \infty}{=}e^{i\eta \ln 
\zeta }\left[ 1+O(\frac{1}{%
i\zeta })\right].  \label{whit1asym1}
\end{eqnarray}
Correspondingly the asymptotic behavior of the Coulomb distorted incident wave
\begin{eqnarray}
\psi_{{\mathbf{k}}}^{(0)}({\mathbf{r}}) \stackrel{r \to \infty}{=}
e^{i\,{\mathbf{k}}\cdot {\mathbf{r}}}e^{i\eta \ln \zeta}\,[1 + 
O(\frac{1}{i\,\zeta})]. \label{asymincwave1}
\end{eqnarray}
The asymptotic behavior of ${\cal{F}}^{(2)}(\zeta)$ is given by
\begin{eqnarray}
{\cal{F}}^{(2)}(i\zeta )\stackrel{\zeta \to 
\infty}{=}\,f^{C}\,\frac{e^{i\zeta }}{r}%
e^{-i\eta \ln 2kr}\left[ 1+O(\frac{1}{i\zeta })\right], 
\label{whit2asymp1}
\end{eqnarray}  
where $f^{C}$ is the on-the-energy-shell Coulomb scattering amplitude:
\begin{eqnarray}
f^{C}=-\eta\,\frac{\Gamma (1+i\eta )}{\Gamma (1-i\eta )}(-i)^{-i\eta
}e^{\pi \eta /2}\frac{e^{-i\eta \ln \sin ^{2}\frac{\theta }{2}}}{2k\sin 
^{2}%
\frac{\theta }{2}}.  \label{scampl1}
\end{eqnarray}
The asymptotic behavior of the scattered wave is given by
\begin{eqnarray}
\psi_{{\mathbf{k}}}^{(sc)}({\mathbf{r}}) \stackrel{r \to \infty}{=}
f^{C}\frac{e^{ikr}}{r}e^{-i\eta \ln 2kr}\,[1 + O(\frac{1}{%
i\zeta })].               \label{asymscwave1}
\end{eqnarray}
Taking into account Eqs. (\ref{twbclscwf2}), (\ref{asymincwave1}), 
(\ref{asymscwave1}) we get the asymptotic behavior 
of the Coulomb scattering wave function for a system of two particles in 
the coordinate space:
\begin{eqnarray}
\psi_{{\mathbf{k}}}^{(+)}({\mathbf{r}}) \stackrel{r \to \infty}{=}
e^{i\,{\mathbf{k}}\cdot {\mathbf{r}}}e^{i\eta \ln \zeta}\,[1 + 
O(\frac{1}{%
i\zeta })]
+f^{C}\frac{e^{ikr}}{r}e^{-i\eta \ln 2kr}\,[1 + O(\frac{1}{%
i\zeta })].               \label{asexpwf1}
\end{eqnarray}
Note that this asymptotic behavior is valid only for 
$|\zeta| \to \infty$. 
For $r \to \infty$ it is valid for all directions in the configuration 
space except for the so-called singular direction, for which 
$\hat {\mathbf{k}} \cdot \hat {\mathbf{r}} = 1$. 
Here $\hat {\mathbf{a}}= {\mathbf{a}}/a$.

One can observe a very interesting feature in the case of the two-body 
Coulomb scattering. The regular solution, according to Eq. (\ref{twbclscwf2}),
consists of two singular solutions, incident and scattered wave, each of them 
also satisfies the Schr\"odinger equation.
Correspondingly, the asymptotic Coulomb scattering wave function 
consists of two terms. The first one, $e^{i{\mathbf{k}}\cdot {\mathbf{r}}}e^{i\eta \ln\zeta}\,(1 + O(\frac{1}{%
i\zeta })$ is the asymptotic form of $e^{i{\mathbf{k}}\cdot 
r}{\cal{F}}^{(1)}(i\zeta)$ and
represents the Coulomb distorted incident wave. The Coulomb distortion 
not only generates
a logarithmic phase factor $\eta\,\ln \zeta$ as an additional phase 
factor 
to the plane wave phase factor ${\mathbf{k}}\cdot {\mathbf{r}}$, but it
also generates an infinite series in powers of $1/\zeta$. This is in 
contrast 
to the two-body scattering problem for particles interacting via  
short-range potentials, where the incident wave is given just by the 
plane wave. The second term in Eq. (\ref{asexpwf1}) is the asymptotic 
form 
for $e^{i{\mathbf{k}}\cdot r}{\cal{F}}^{(2)}(i\zeta)$. It generates the 
outgoing two-body spherical wave times the Coulomb scattering amplitude and also contains a factor which can be written as an asymptotic 
expansion in powers of $1/\zeta$. 

\section{Asymptotic three-body incident wave}

After consideration of the incident wave for the two-body case, it is 
easier to proceed to the incident wave for the three-body case. 
By definition, the incident three-body wave is the part of 
the total three-body scattering wave function of the first type, which 
does not contain two- and three-body scattered waves.
We have shown that the two-body incident wave is a singular solution of 
the two-body \SE. It is naturally to ask whether the three-body incident wave is a solution of the three-body \SE. An educated guess tells us that the answer may be "yes". 
First, the scattering wave function of the first type which consists of the three-body incident and scattered waves (two- and three-body) is a regular solution of the three-body \SE. However there are also singular solutions of the three-body \SE.
And it is very plausible that the three-body incident wave is one of the singular solutions while the scattered wave represents another singular solution. However, we cannot prove it until an analytical expression for the three-body incident wave will be available.  

Our goal in this work is to derive the asymptotic incident three-body wave function in the leading orders $O(1),\,O(1/\rho_{\alpha}^{2}),\,O(1/\rho_{\alpha}^{2})$ in the asymptotic region $\Omega_{\alpha}$, where particles $\beta$ and $\gamma$ are close to each other and far away from particle $\alpha$. We will demonstrate that the terms of order $O(1/{\rho}_{\alpha}^{2})$ can be derived without explicit solution of the three-body \SE. 
In principle the method we use can be applied to get even the higher order terms 
of the three-body incident wave but it is worth mentioning that the next 
order term in the asymptotic expansion of the three-body incident wave 
$O(1/{\rho}_{\alpha}^{3})$ is inferior to the outgoing three-body scattered wave 
$O(1/R^{5/2})$. Hence for practical applications one need to know the three-body
scattered wave before getting the term $O(1/{\rho}_{\alpha}^{3})$ in the asymptotic 
expansion of the three-body incident wave. A knowledge of the three-body incident wave up to terms $O(1/{\rho}_{\alpha}^{3})$ allows us to write down the leading asymptotic terms of the three-body scattering wave function of the first type in the asymptotic region $\Omega_{\alpha}$ up to terms $O(1/{\rho}_{\alpha}^{3})$. 
Note that the expressions for the asymptotic incident 
three-body wave in two other asymptotic regions 
$\Omega_{\beta}$ and $\Omega_{\gamma}$ can be derived by simple cyclic 
permutation of indexes $\alpha,\,\beta$ and $\gamma$. As we have mentioned 
earlier, the asymptotic incident three-body wave is the part of 
the total three-body scattering wave function of the first type, which 
does not contain two- and three-body scattered waves. This wave function 
should smoothly transform into the asymptotic incident three-body wave 
function in the asymptotic region $\Omega_{0}$. This smooth matching is 
an important part of the boundary conditions that provides a unique solution. 

The leading asymptotic term of the three-body incident wave in 
$\Omega_{0}$ derived by Redmond \cite{Redmond,R73} is given by 
\begin{equation}
\Psi_{{\mathrm{\mathbf{k}}}_{\alpha},{\mathrm{\mathbf{q}}}_{\alpha}}^{(0)(+)}(\mathrm{\mathbf{r}}_{\alpha},{\brho}_{\alpha})=
e^{i\,\mathrm{\mathbf{k}}_{\alpha}\cdot \mathrm{\mathbf{r}}_{\alpha }}\,
{e^{ i\,{\mathrm{\mathbf{q}}}_{\alpha} \cdot {\mathbf{\brho}}_{\alpha 
} }}\,\prod\limits_{\nu=\alpha,\,\beta,\,\gamma}\,e^{i\,\eta_{\nu}\ln\,\zeta_{\nu}},  
\label{redm72}
\end{equation}
where
\be
\zeta_{\nu} = k_{\nu}\,r_{\nu} - {\rk}_{\nu}\cdot {\rm {\bf r}}_{\nu}. \label{zetanu1}
\ee 
\be
\eta_{\alpha}= \frac{Z_{\beta}\,Z_{\gamma}\,e^{2}\,\mu_{\alpha}}{k_{\alpha}} 
\label{coulombpar1}
\ee
is the Coulomb parameter of particles $\beta$ and $\gamma$, $\mu_{\alpha}$ is the reduced mass of 
particles $\beta$ and $\gamma$.
It is the three-body Coulomb distorted plane wave. For practical 
applications 
Merkuriev \cite{FM93}, 
Garibotti and Miraglia \cite{GM80} extended the asymptotic Redmond's term 
by substituting the confluent hypergeometric functions for the 
exponential 
Coulomb distortion factors. This extended wave function, often called 
the 3C wave 
function, is given by  
\begin{equation}
\Psi_{{\mathrm{\mathbf{k}}}_{\alpha},{\mathrm{\mathbf{q}}}_{\alpha}}^{(3C)(+)}(\mathrm{\mathbf{r}}_{\alpha},{\brho}_{\alpha 
})=e^{i\,\mathrm{\mathbf{k}}_{\alpha}\cdot \mathrm{\mathbf{r}}_{\alpha }}\,
e^{i\,{\mathrm{\mathbf{q}}}_{\alpha}\cdot {\brho}_{\alpha}}\,\prod\limits_{\nu=\alpha,\,\beta,\,\gamma}\,F_{\nu}(\zeta_{\nu}),  \label{redmer1}
\end{equation}
where
\be
F_{\nu}(\zeta_{\nu})= N_{\nu}\,{}_{1}F_{1}(-i\,\eta_{\nu},\,1 ;i\,\zeta_{\nu}), 
\label{hypergeom1}
\ee
${}_{1}F_{1}(-i\,\eta_{\nu},\,1 ;i\,\zeta_{\nu})$ is the confluent hypergeometric 
function and 
\be
N_{\nu}= e^{-\pi\,\eta_{\nu}/2}\,\Gamma(1+ i\,\eta_{\nu}). 
\label{normfactr1}
\ee 
Note that
\be
\psi_{\mathrm{\mathbf{k}}_{\alpha}}(\mathrm{\mathbf{r}}_{\alpha })=
e^{i\,\mathrm{\mathbf{k}}_{\alpha}\cdot \mathrm{\mathbf{r}}_{\alpha }}\,
F_{\nu}(\zeta_{\nu})  \label{coulscatwf1}
\ee
is the Coulomb scattering wave function of particles $\beta$ and $\gamma$ moving with
the relative momentum $\mathrm{\mathbf{k}}_{\alpha}$ 
and is well-behaved even in 
the singular 
directions ($\zeta_{\nu}< C $ for $r_{\nu} \to \infty$) where the 
Redmond's 
asymptotic term is not determined. If any of the particles is neutral, 
then 
the resulting asymptotic solution becomes the plane wave for the neutral 
particle and the exact two-body scattering wave function for the charged 
pair. However, neither Redmond's asymptotic term
$\Psi_{{\mathrm{\mathbf{k}}}_{\alpha},{\mathrm{\mathbf{q}}}_{\alpha}}^{(0
)(+)}
(\mathrm{\mathbf{r}}_{\alpha},{\brho}_{\alpha})$ nor 
the 3C wave function 
$\Psi_{{\mathrm{\mathbf{k}}}_{\alpha},{\mathrm{\mathbf{q}}}_{\alpha}}^{(3
C)(+)}
(\mathrm{\mathbf{r}}_{\alpha},{\brho}_{\alpha})$ are 
asymptotic solutions of the \SE in the asymptotic 
domains $\Omega_{\nu}$, $\,\nu=\alpha,\beta,\gamma$. Redmond's 
asymptotic term, by construction, satisfies the asymptotic \SE up to terms  
$O(1/r_{\alpha}^{2},\,1/r_{\beta}^{2},\,1/r_{\gamma}^{2})$. However, in 
the asymptotic region, $\Omega_{\nu}$, the distance between the 
particles of pair $\nu$ is limited: $r_{\nu} < C'$. Hence the terms 
$O(1/r_{\nu})$ are not small and the potential $V^{C}_{\nu}$ in the \SE 
has to be compensated exactly rather 
than asymptotically as happens when we use the Redmond's asymptotic 
wave function in $\Omega_{0}$. In the 3C wave function two very 
important effects are absent. Consider, for example, the asymptotic 
region $\Omega_{\alpha}$. In this region $r_{\alpha} << \rho_{\alpha}$. 
Hence the two-body relative motion of particles $\beta$ and $\gamma$ is 
distorted by the Coulomb field of the third particle $\alpha$ \cite{AM93}. 
The second evident defect in the 3C function is the absence of the nuclear 
interaction between particles $\beta$ and $\gamma$ which can be close 
enough to each other in $\Omega_{\alpha}$. Nevertheless, the 3C wave 
function can be used as a starting point to derive the 
the leading asymptotic terms of the three-body incident wave in 
$\Omega_{\alpha}$ \cite{AM93,ML96}, because this asymptotic three-body 
incident wave should match the Redmond's asymptotic term in $\Omega_{0}$. 
We will demonstrate now how important the condition of the matching of the 
asymptotic wave functions is on the border of different asymptotic 
regions \cite{AM93}. 

In the Redmond's asymptotic incident wave three logarithmic phase 
factors appear, one phase factor for each pair rather than two phase factors 
in the factorized solution. 
It is a very important conclusion. In all the conventional approaches 
for breakup processes, including coupled channels codes like FRESCO, the 
three-body scattering wave function is approximated by the factorized 
one. From the consideration above, it is clear that if Coulomb 
interactions are important, such an approximation is not accurate. If 
the interactions are short-range, the factorized solution matches the 
asymptotic solution in $\Omega_{0}$ and is justified in the asymptotic 
region $\Omega_{\alpha}$. It has been shown in \cite{AM93,ML96} 
that the actual asymptotic solution of the asymptotic \SE
$\Psi_{{\mathbf{k}}_{\alpha}{\mathbf{q}}_{\alpha}}^{(as)}
({\mathbf{r}}_{\alpha},{\brho}_{\alpha})$, which matches the Redmond's 
asymptotic term in $\Omega_{0}$, cannot be written in a factorized form 
and has a quite complicated behavior.
In \cite{AM93,ML96}  all the leading asymptotic terms up to  
$O(1/{\rho}_{\alpha}^{2})$ of the asymptotic wave function 
$\Psi_{{\mathbf{k}}_{\alpha}{\mathbf{q}}_{\alpha}}^{(as)}
({\mathbf{r}}_{\alpha},{\brho}_{\alpha})$ have been derived in the 
asymptotic region $\Omega_{\alpha}$. In this work we will 
present a derivation of the expansion of the asymptotic wave function, 
$\Psi_{{\mathbf{k}}_{\alpha}{\mathbf{q}}_{\alpha}}^{(as)}
({\mathbf{r}}_{\alpha},{\brho}_{\alpha})$, up to terms 
$O(1/{\rho}_{\alpha}^{3})$.
The derived asymptotic expansion contains all the terms  
$O(1),\,O(1/{\rho}_{\alpha})$ and $O(1/{\rho}_{\alpha}^{2})$.   
Since we are looking for the terms $O(1/{\rho}_{\alpha}^{2})$, 
we need to keep the terms up to $O(1/{\rho}_{\alpha}^{3})$. 
Instead of the asymptotic expansion of the Coulomb potentials 
$V^{C}_{\beta}({\rm {\bf r}}_{\beta})$ and $V^{C}_{\gamma}({\rm {\bf 
r}}_{\gamma})$ in terms of $1/{\rho}_{\alpha}$, we will start our 
derivation from the exact three-body \SE (\ref{ThSE}).
The terms of $O(1/{\rho}_{\alpha}^{3})$ will be dropped later.
The asymptotic wave function in $\Omega_{\alpha}$ should match the 
asymptotic wave function in $\Omega_{0}$.  
The 3C wave function satisfies Eq.(\ref{ThSE}) 
up to terms $O(1/r_{\alpha}^{2},1/{\rho}_{\alpha}^{2})$ and we 
can use it as the initial wave function. However, this wave function 
should be modified to satisfy the \SE in $\Omega_{\alpha}$.   
Note that usually in the literature it is assumed that the Redmond's 
asymptotic term satisfies the \SE in $\Omega_{0}$ in the leading order only.
First we will show that the 3C wave function satisfies the \SE in 
$\Omega_{0}$ up terms of order $O(1/r_{\nu}^{2})$. 
To this end we just substitute the 3C wave function (\ref{redmer1}) 
into the \SE (\ref{ThSE}):
\begin{eqnarray}
&&(E-T_{\mathbf{r}_{\alpha }}-T_{{\brho}_{\alpha }}-V)[e^{i\mathbf{k}%
_{\alpha }\cdot \mathbf{r}_{\alpha }+i\mathbf{q}_{\alpha }\cdot {\brho}%
_{\alpha }}\,\varphi _{{\mathbf{k}}_{\alpha }}(\mathbf{r}_{\alpha 
})\varphi
_{{\mathbf{k}}_{\beta }}(\mathbf{r}_{\beta })\varphi 
_{{\mathbf{k}}_{\gamma
}}(\mathbf{r}_{\gamma })]  \notag \\
&&=e^{i\mathbf{k}_{\alpha }\cdot \mathbf{r}_{\alpha 
}+i\mathbf{q}_{\alpha
}\cdot {\brho}_{\alpha }}\varphi _{{\mathbf{k}}_{\beta 
}}(\mathbf{r}_{\beta
})\varphi _{{\mathbf{k}}_{\gamma }}(\mathbf{r}_{\gamma })[\frac{%
\bigtriangleup _{\mathbf{r}_{\alpha }}}{2\mu _{\alpha 
}}+\frac{i{\mathbf{k}}%
_{\alpha }\cdot \bigtriangledown _{\mathbf{r}_{\alpha }}}{\mu _{\alpha 
}}%
-V_{\alpha } \nonumber\\
&&+\frac{\bigtriangleup _{{\brho}_{\alpha }}}{2M_{\alpha 
}}+\frac{i[\mathbf{q%
}_{\alpha }-i\sum\limits_{\nu =\beta ,\gamma }\bigtriangledown 
_{{\brho}%
_{\alpha }}\ln \varphi _{{\mathbf{k}}_{\nu }}]\cdot \bigtriangledown 
_{{\brho%
}_{\alpha }}}{M_{\alpha }}  \nonumber\\
&&+\frac{\bigtriangledown _{\mathbf{r}_{\alpha }}\varphi _{{\mathbf{k}}%
_{\gamma }}\cdot \bigtriangledown _{\mathbf{r}_{\alpha }}\varphi 
_{{\mathbf{k%
}}_{\beta }}}{\mu _{\alpha }\varphi _{{\mathbf{k}}_{\beta }}\varphi _{{%
\mathbf{k}}_{\gamma }}}+\frac{\bigtriangledown _{{\brho}_{\alpha 
}}\varphi _{%
{\mathbf{k}}_{\gamma }}\cdot \bigtriangledown_{\brho_{\alpha}}
\varphi _{{\mathbf{k}}_{\beta }}}{M_{\alpha }\varphi 
_{{\mathbf{k}}_{\beta
}}\varphi _{{\mathbf{k}}_{\gamma }}}]\varphi _{{\mathbf{k}}_{\alpha }}(%
\mathbf{r}_{\alpha })  \nonumber\\
&&+e^{i\mathbf{k}_{\beta }\cdot \mathbf{r}_{\beta }+i\mathbf{q}_{\beta
}\cdot {\brho}_{\beta }}\varphi _{{\mathbf{k}}_{\alpha 
}}(\mathbf{r}_{\alpha
})\varphi _{{\mathbf{k}}_{\gamma }}(\mathbf{r}_{\gamma })[\frac{%
\bigtriangleup _{\mathbf{r}_{\beta }}}{2\mu _{\beta 
}}+\frac{i{\mathbf{k}}%
_{\beta }\cdot \bigtriangledown _{\mathbf{r}_{\beta }}}{\mu _{\beta }}%
-V_{\beta }  \nonumber\\
&&+\frac{\bigtriangleup _{{\brho}_{\beta }}}{2M_{\beta 
}}+\frac{i[\mathbf{q}%
_{\beta }-i\sum\limits_{\tau =\alpha ,\gamma }\bigtriangledown 
_{{\brho}%
_{\beta }}\ln \varphi _{{\mathbf{k}}_{\tau }}]\cdot \bigtriangledown 
_{{\brho%
}_{\beta }}}{M_{\beta }} \nonumber\\
&&+\frac{\bigtriangledown _{\mathbf{r}_{\beta }}\varphi _{{\mathbf{k}}%
_{\gamma }}\cdot \bigtriangledown _{\mathbf{r}_{\beta }}\varphi _{{k}%
_{\alpha }}}{\mu _{\beta }\varphi _{{\mathbf{k}}_{\alpha }}\varphi _{{%
\mathbf{k}}_{\gamma }}}+\frac{\bigtriangledown _{{\brho}_{\beta 
}}\varphi _{{%
\mathbf{k}}_{\gamma }}\cdot \bigtriangledown _{{\brho}_{\beta }}\varphi 
_{{%
\mathbf{k}}_{\alpha }}}{M_{\beta }\varphi _{{\mathbf{k}}_{\alpha 
}}\varphi _{%
{\mathbf{k}}_{\gamma }}}]\varphi _{{\mathbf{k}}_{\beta 
}}(\mathbf{r}_{\beta
}) \nonumber\\
&&+e^{i\mathbf{k}_{\gamma }\cdot \mathbf{r}_{\gamma 
}+i\mathbf{q}_{\gamma
}\cdot {\brho}_{\gamma }}\varphi _{{\mathbf{k}}_{\alpha }}(\mathbf{r}%
_{\alpha })\varphi _{{\mathbf{k}}_{\beta }}(\mathbf{r}_{\beta })[\frac{%
\bigtriangleup _{\mathbf{r}_{\gamma }}}{2\mu _{\gamma 
}}+\frac{i{\mathbf{k}}%
_{\gamma }\cdot \bigtriangledown _{\mathbf{r}_{\gamma }}}{\mu _{\gamma 
}}%
-V_{\gamma }  \nonumber\\
&&+\frac{\bigtriangleup _{{\brho}_{\gamma }}}{2M_{\gamma 
}}+\frac{i[\mathbf{q%
}_{\gamma }-i\sum\limits_{\omega =\alpha ,\beta }\bigtriangledown 
_{{\brho}%
_{\gamma }}\ln \varphi _{{\mathbf{k}}_{\omega }}]\cdot 
\bigtriangledown_{\brho_{\gamma }}}{M_{\gamma }}  \label{omega0maineqn1} 
\nonumber\\
&&+\frac{\bigtriangledown _{\mathbf{r}_{\gamma }}\varphi _{{\mathbf{k}}%
_{\beta }}\cdot \bigtriangledown _{\mathbf{r}_{\gamma }}\varphi _{{k}%
_{\alpha }}}{\mu _{\beta }\varphi _{{\mathbf{k}}_{\alpha }}\varphi _{{%
\mathbf{k}}_{\beta }}}+\frac{\bigtriangledown _{{\brho}_{\gamma 
}}\varphi _{{%
\mathbf{k}}_{\beta }}\cdot \bigtriangledown _{{\brho}_{\gamma }}\varphi 
_{{%
\mathbf{k}}_{\alpha }}}{M_{\gamma }\varphi _{{\mathbf{k}}_{\alpha 
}}\varphi
_{\widetilde{\mathbf{k}}_{\beta }}}]\varphi _{{\mathbf{k}}_{\gamma }}
(\mathbf{r}_{\gamma }).  
\label{omega0maineqn2}
\end{eqnarray}
Here
\begin{eqnarray}
\varphi_{{\mathbf{k}}_{\nu}}({\mathbf{r}}_{\nu})=N_{\nu}\,
F(-i\eta_{\nu},\,1 ; i\,\zeta_{\nu})= {\mathcal F}_{\nu}^{(1)}(i\,\zeta_{\nu}) + 
{\mathcal F}_{\nu}^{(2)}(i\,\zeta_{\nu}). \label{varphi12}
\end{eqnarray}
Taking into account 
\begin{eqnarray}
\lbrack \frac{\bigtriangleup _{{\mathbf{r}}_{\nu }}}{2\mu
_{\nu }}+\frac{i{\mathbf{k}}_{\nu }\cdot \bigtriangledown _{%
{\mathbf{r}}_{\nu }}}{\mu _{\nu }}-V_{\nu }]\varphi _{{\mathbf{k}}_{\nu 
}}({\mathbf{r}}_{\nu})=0
\end{eqnarray}
we derive
\begin{eqnarray}
(E-T_{\mathbf{r}_{\alpha }}-T_{{\brho}_{\alpha
}}-V)[e^{i\mathbf{k}_{\alpha }\cdot \mathbf{r}_{\alpha }+i%
\mathbf{q}_{\alpha }\cdot {\brho}_{\alpha }}\,\varphi
_{{\mathbf{k}}_{\alpha }}(\mathbf{r}_{\alpha })
\varphi_{\mathbf{k}_{\beta }}(\mathbf{r}_{\beta })\varphi 
_{{\mathbf{k}}_{\gamma }}(%
\mathbf{r}_{\gamma})]\nonumber\\=O(1/r_{\alpha}^{2},1/r_{\beta}^{2},1/r
_{\gamma}^{2}).
\label{sheqin0reg}
\end{eqnarray}
We did not use any approximation to get Eq. \ref{sheqin0reg}. 
Thus the 3C wave function indeed satisfies the \SE in $\Omega_{0}$ up to 
the terms $O(1/r_{\alpha}^{2},1/r_{\beta}^{2},1/r_{\gamma}^{2})$,
i. e. after substitution of the 3C wave function into the \SE all the terms of order
$O(1)$ and $O(1/r_{\alpha})$ are exactly compensated. 
Hence the 3C wave function can be used as a starting wave function with a proper 
modifications to look for an asymptotic solution in $\Omega_{\alpha}$. Taking into account Eq. (\ref{varphi12}) we can rewrite the 3C wave function as 
in a form which is suitable for 
consideration in the $\Omega_{\alpha}$ asymptotic domain:    
\begin{eqnarray}
\Psi^{(3C)(+)}_{{\rm{\bf{k}}}_{\alpha},{\rm{\bf{q}}}_{\alpha}}
({\rm{\bf{r}}}_{\alpha},{\brho}_{\alpha})\,=
e^{i\,{\rm{\bf{k}}}_{\alpha}{\cdot}{\rm{\bf{r}}}_{\alpha}}\,
e^{i\,{\rm{\bf{q}}}_{\alpha}{\cdot}{\brho}_{\alpha}}\nonumber\\
\times\,\lbrack {\cal F}_{\beta}^{(1)}(i\zeta_{\beta})
{\cal F}_{\gamma}^{(1)}(i\zeta_{\gamma}) 
N_{\alpha}F_{\alpha}(i\zeta_{\alpha}) 
+ {\cal F}_{\beta}^{(2)}(i\zeta_{\beta}) 
{\cal F}_{\gamma}^{(1)}(i\zeta_{\gamma})
N_{\alpha}F_{\alpha}(i\zeta_{\alpha})  \\
+{\cal F}_{\beta}^{(1)}(i\zeta_{\beta}) 
{\cal F}_{\gamma}^{(2)}(i\zeta_{\gamma})
N_{\alpha}F_{\alpha}(i\zeta_{\alpha}) 
+ {\cal F}_{\beta}^{(2)}(i\zeta_{\beta}) 
{\cal F}_{\gamma}^{(2)}(i\zeta_{\gamma})
N_{\alpha}F_{\alpha}(i\zeta_{\alpha}) \rbrack. \nonumber
\label{psi1}
\end{eqnarray}
Here, asymptotically, for $|\zeta_{\nu}| \to \infty$, the first term 
${\cal{F}} ^{(1)}(\zeta_{\nu}) \sim O(1)$ and the second term 
${\cal{F}}^{(2)}(\zeta_{\nu}) \sim O(1/\zeta_{\nu})$. Hence in the asymptotic 
domain $\Omega_{\alpha}$ \, ${\cal{F}} ^{(1)}(\zeta_{\nu})$ and 
${\cal{F}}^{(2)}(\zeta_{\nu}),\;\nu=\beta,\,\gamma$,
can be treated asymptotically while 
$\varphi_{{\mathbf{k}}_{\alpha}}({\mathbf{r}}_{\alpha})$
should be considered explicitly, because $\Omega_{\alpha}$ includes the 
region, where $r_{\alpha}$ is limited. 
Moreover, in the asymptotic region $\Omega_{\alpha}$ the relative motion 
of particles 
$\beta$ and $\gamma$ is distorted by the third particle $\alpha$ 
due to the long-range Coulomb interaction. It means that 
the wave function of the relative motion of particles $\beta$ and 
$\gamma$ in $\Omega_{\alpha}$  will be different from
the wave function 
$e^{i\,{\rm{\bf{k}}}_{\alpha}{\cdot}{\rm{\bf{r}}}_{\alpha}}\,
N_{\alpha}F_{\alpha}(i\zeta_{\alpha})$  describing the relative motion 
of particles $\beta$ and $\gamma$ in the absence of the third particle. 
Since interacting particles $\beta$ and $\gamma$ can be close to each 
other in $\Omega_{\alpha}$, their nuclear interaction should also be taken
into account. Following \cite{ML96} we replace each 
$N_{\alpha}\,F_{\alpha}(i\zeta_{\alpha})$ in Eq. (\ref{psi1}) by the corresponding 
unknown function 
${\varphi}_{\alpha}^{(nm)}({\rm{\bf{r}}}_{\alpha}),\,n,m=1,2$: 
\begin{eqnarray}
\Psi^{{(as)(+)}}_{{\rm{\bf{k}}}_{\alpha},{\rm{\bf{q}}}_{\alpha}}
({\rm{\bf{r}}}_{\alpha},{\brho}_{\alpha})\,=
e^{i\,{\rm{\bf{k}}}_{\alpha}{\cdot}{\rm{\bf{r}}}_{\alpha}}\,
e^{i\,{\rm{\bf{q}}}_{\alpha}{\cdot}{\brho}_{\alpha}}\,\nonumber \\
\times\lbrack {\cal F}_{\beta}^{(1)}(i\zeta_{\beta}) 
{\cal F}_{\gamma}^{(1)}(i\zeta_{\gamma})
{\varphi}_{\alpha}^{(11)}({\rm{\bf{r}}}_{\alpha},\,
{\brho}_{\alpha}) + 
{\cal F}_{\beta}^{(2)}(i\zeta_{\beta}) 
{\cal F}_{\gamma}^{(1)}(i\zeta_{\gamma})
{\varphi}_{\alpha}^{(21)}({\rm{\bf{r}}}_{\alpha},\,
{\brho}_{\alpha}) \nonumber\\ 
+{\cal F}_{\beta}^{(1)}(i\zeta_{\beta}) 
{\cal F}_{\gamma}^{(2)}(i\zeta_{\gamma})
{\varphi}_{\alpha}^{(12)}({\rm{\bf{r}}}_{\alpha},\,
{\brho}_{\alpha}) + 
{\cal F}_{\beta}^{(2)}(i\zeta_{\beta}) 
{\cal F}_{\gamma}^{(2)}(i\zeta_{\gamma})
{\varphi}_{\alpha}^{(22)}({\rm{\bf{r}}}_{\alpha},\,
{\brho}_{\alpha})\rbrack. 
\label{psi2}
\end{eqnarray}
Derivation of 
${\varphi}_{\alpha}^{(nm)}({\rm{\bf{r}}}_{\alpha}),\,n,m=1,2$ is our 
final goal. 
Now we substitute Eq. (\ref{psi2}) into the {\SE\;} (\ref{ThSE}).
When substituting Eq. (\ref{psi2}) into the {\SE\,} we assume that each 
term of the sum (\ref{psi2}) satisfies the \SE. Moreover, as we will see, each function ${\varphi}_{\alpha}({\rm{\bf{r}}}_{\alpha},\,{\brho}_{\alpha})$ depends 
on the preceding functions ${\cal F}_{\beta}^{(n)}(i\zeta_{\beta})
{\cal F}_{\gamma}^{(m)}(i\zeta_{\gamma})$ where $n,m=1,2$,
i.e. for each term in (\ref{psi2}) the modification 
is different.
We also take into account that 
\begin{equation}
({\frac{1}{2 
\mu_{\nu}}}{\mbox{\boldmath{$\Delta$}}}_{{\rm{\bf{r}}}_{\nu}} +
i\,{\frac{1}{\mu_{\nu}}}    
\,{\rm{\bf{k}}}_{\nu}{\cdot}
{\mbox{\boldmath{$\nabla$}}}_{{\rm{\bf{r}}}_{\nu}} -
V_{\nu}^{C})\,{\cal 
F}_{\nu}^{(1,2)}(i\zeta_{\nu})=0. 
\label{hyp1}
\end{equation}
Substitution of the first term of Eq. (\ref{psi2}) into the {\SE\,} 
generates the equation for
${\varphi}_{\alpha}^{(11)}({\rm{\bf{r}}}_{\alpha})$:
\begin{eqnarray}
{\cal F}^{(1)}_{\beta}(i\zeta_{\beta})
{\cal F}^{(1)}_{\gamma}(i\zeta_{\gamma})
\lbrack{\frac{1}{2\mu_{\alpha}}}{{\mbox{\boldmath{$\Delta$}}}_{{\rm{\bf{r
}}}_{\alpha}}} +
{\frac{1}{2 
M_{\alpha}}}{{\mbox{\boldmath{$\Delta$}}}_{{\brho}_{\alpha}}} +
i\,{\frac{1}{\mu_{\alpha}}}    
\,{\rm{\bf{k}}}_{\alpha}{\cdot}
{\mbox{\boldmath{$\nabla$}}}_{{\rm{\bf{r}}}_{\alpha}} +
i\,{\frac{1}{M_{\alpha}}} \,
{\rm{\bf{q}}}_{\alpha}{\cdot}
{\mbox{\boldmath{$\nabla$}}}_{{\brho}_{\alpha}} +
\nonumber \\
{\frac{1}{\mu_{\alpha}}}\,{\sum_{\nu=\beta,\gamma}} 
{\mbox{\boldmath{$\nabla$}}}_{{\rm{\bf{r}}}_{\alpha}}
\ln {\cal F}^{(1)}_{\nu}(i\zeta_{\nu}){\cdot}
{\mbox{\boldmath{$\nabla$}}}_{{\rm{\bf{r}}}_{\alpha}} +
{\frac{1}{M_{\alpha}}}\,{\sum_{\nu=\beta,\gamma}} 
{\mbox{\boldmath{$\nabla$}}}_{{\brho}_{\alpha}}
\ln {\cal F}^{(1)}_{\nu}(i\zeta_{\nu}){\cdot}
{\mbox{\boldmath{$\nabla$}}}_{{\brho}_{\alpha}} - 
V_{\alpha}({\rm{\bf{r}}}_{\alpha}) + \nonumber \\
{\frac{1}{\mu_{\alpha}}}\, 
{\mbox{\boldmath{$\nabla$}}}_{{\rm{\bf{r}}}_{\alpha}}
\ln {\cal F}^{(1)}_{\beta}(i\zeta_{\beta}){\cdot}
{\mbox{\boldmath{$\nabla$}}}_{{\rm{\bf{r}}}_{\alpha}}
\ln {\cal F}^{(1)}_{\gamma}(i\zeta_{\gamma}) + 
{\frac{1}{M_{\alpha}}}\,
{\mbox{\boldmath{$\nabla$}}}_{{\brho}_{\alpha}}
\ln {\cal F}^{(1)}_{\beta}(i\zeta_{\beta}){\cdot}
{\mbox{\boldmath{$\nabla$}}}_{{\brho}_{\alpha}}
\ln {\cal F}^{(1)}_{\gamma}(i\zeta_{\gamma})\rbrack \nonumber \\
\times \varphi_{\alpha}^{(11)}({\rm{\bf{r}}}_{\alpha},\,
{\brho}_{\alpha})=0. \label{psi3}
\end{eqnarray}
Since particles $\beta$ and $\gamma$ are allowed to be close in 
$\Omega_{\alpha}$ 
their interaction potential is given by the sum of the
Coulomb and nuclear potentials.
Now we will simplify this equation by dropping all the terms  
$O(1/{\rho}_{\alpha}^{3})$ and explicitly compensate all the terms 
$O(1),\,O(1/{\rho}_{\alpha}),\,O(1/{\rho}_{\alpha}^{2})$. We consider 
only the 
nonsingular directions, i. e. ${\rm{\bf{\hat k}}}_{\nu}{\cdot}{\rm {\bf 
{\hat r}}}_{\nu} \not=1,\; \nu=\beta, \gamma$. 
To analyze the fifth term in the brackets we use equations
\begin{equation}
{\cal F}_{\nu}^{(1)}(i\zeta_{\nu})=
{\tilde{\cal F}}_{\nu}^{(1)}(i\zeta_{\nu})\lbrack 1 - 
i\,\frac{\eta_{\nu}^{2}}
{\zeta_{\nu}}+  O(1/\zeta_{\nu}^{2}) \rbrack, \label{fas1} 
\end{equation}
\begin{equation}
{\tilde{\cal 
F}}_{\nu}^{(1)}(i\zeta_{\nu})=e^{i\,{\eta}_{\nu}\ln{\zeta_{\nu}}},
\label{fntld1}
\end{equation}
\begin{equation}
{\mbox{\boldmath{$\nabla$}}}_{{\rm{\bf{r}}}_{\alpha}} 
\ln {\cal F}^{(1)}_{\nu}(i\zeta_{\nu})= 
{\mbox{\boldmath{$\nabla$}}}_{{\rm{\bf{r}}}_{\alpha}} 
\ln {\tilde {\cal F}}^{(1)}_{\nu}(i\zeta_{\nu}) - 
i\frac{m_{\nu}}{m_{\beta \gamma}} 
\frac{\eta_{\nu}^{2}}{k_{\nu}\,r_{\nu}^{2}}\,\frac{ {\hat {\rm {\bf 
r}}}_{\nu} - {\hat {\rm {\bf k}}}_{\nu} }{(1 - {\hat {\rm {\bf 
k}}}_{\nu}\cdot {\hat {\rm {\bf r}}}_{\nu})^{2}} + O(1/r_{\nu}^{3}) ,  
\label{tldt1} 
\end{equation}
\begin{eqnarray}
&&{\mbox{\boldmath{$\nabla$}}}_{\mathrm{\mathbf{r}}_{\alpha }}\ln 
{\tilde{%
\mathcal{F}}}_{\nu }^{(1)}(i\zeta _{\nu }) 
={\mbox{\boldmath{$\nabla$}}}_{%
\mathrm{\mathbf{r}}_{\alpha }}e^{\frac{m_{\nu }}{m_{\beta \gamma 
}}\epsilon
_{\nu \,\alpha }\mathrm{\mathbf{r}_{\alpha }\cdot }{\mbox{\boldmath{$%
\nabla$}}}_{{\brho}_{\alpha }}}\ln {\tilde{\mathcal{F}}}_{\nu 
}^{(1)}(i\zeta
_{\nu \alpha }) \nonumber\\
&&=\epsilon _{\nu \,\alpha }\,\frac{m_{\nu }}{m_{{\beta }{\gamma 
}}}\,{%
\mbox{\boldmath{$\nabla$}}}_{{\brho}_{\alpha }}\ln 
{\tilde{\mathcal{F}}}%
_{\nu }^{(1)}(i\zeta _{\nu \,\alpha })+\frac{m_{\nu }^{2}}{m_{\beta 
\,\gamma
}^{2}}(\mathrm{\mathbf{r}_{\alpha }\cdot 
\,{\mbox{\boldmath{$\nabla$}}}_{{%
\brho}_{\alpha }})\,\mbox{\boldmath{$\nabla$}}}_{{\brho}_{\alpha }}\ln 
{%
\tilde{\mathcal{F}}}_{\nu }^{(1)}(i\zeta _{\nu \,\alpha }),
\label{tfnu1}
\end{eqnarray}
\begin{eqnarray}
{\mbox{\boldmath{$\nabla$}}}_{{\brho}_{\alpha}} 
\ln {\tilde {\cal F}}^{(1)}_{\nu}(i\zeta_{\nu\,\alpha})=
i\eta_{\nu}\,\epsilon_{\nu \alpha}\,
\frac{1}{\rho_{\alpha}}
\frac{ {\hat{\rm {\bf k}}}_{\nu} - \epsilon_{\alpha \nu}
{\hat {\brho}}_{\alpha}}
{1 -\epsilon_{\alpha \nu}{\hat{\rm {\bf k}}}_{\nu}\cdot
{\hat {\brho}}_{\alpha}} +
O(\frac{1}{\rho_{\alpha}^{2}}),    \label{sss1} 
\end{eqnarray}
\begin{equation}
{\mbox{\boldmath{$\nabla$}}}_{{\rm{\bf{r}}}_{\alpha}}\lbrack 
-i\,\frac{\eta_{\nu}^{2}}
{\zeta_{\nu}} \rbrack = 
i\,\eta_{\nu}^{2}\,\frac{m_{\nu}}{m_{\beta\,\gamma}}\,
\frac{1}{k_{\nu}\,\rho_{\alpha}^{2}}\,
\frac{ {\hat{\rm {\bf k}}}_{\nu} - \epsilon_{\alpha \nu}
{\hat {\brho}}_{\alpha}}
{(1 -\epsilon_{\alpha \nu}{\hat{\rm {\bf k}}}_{\nu}\cdot
{\hat {\brho}}_{\alpha})^{2}} +
O(\frac{1}{\rho_{\alpha}^{3}}). \label{ss1} 
\end{equation}
To estimate the sixth and the ninth terms we use equations
\begin{eqnarray}
{\mbox{\boldmath{$\nabla$}}}_{{\brho}_{\alpha}} 
\ln {\cal F}^{(1)}_{\nu}(i\zeta_{\nu})=
i\eta_{\nu}\frac{1}{r_{\nu}}\,\epsilon_{\nu \alpha}\,
\frac{ {\hat{\rm {\bf k}}}_{\nu} -{\hat {\rm {\bf r}}}_{\nu}}
{1 - {\hat{\rm {\bf k}}}_{\nu}\cdot{\hat {\rm {\bf r}}}_{\nu}} +
O(\frac{1}{r_{\nu}^{2}}) \label{as2} \\
=i\eta_{\nu}\,\epsilon_{\nu \alpha}\,
\frac{1}{\rho_{\alpha}}
\frac{ {\hat{\rm {\bf k}}}_{\nu} - \epsilon_{\alpha \nu}
{\hat {\brho}}_{\alpha}}
{1 -\epsilon_{\alpha \nu}{\hat{\rm {\bf k}}}_{\nu}\cdot
{\hat {\brho}}_{\alpha}} +
O(\frac{1}{\rho_{\alpha}^{2}}). \label{ss2} 
\end{eqnarray}
To estimate the eigth term we use equation
\begin{equation}
{\mbox{\boldmath{$\nabla$}}}_{{\rm{\bf{r}}}_{\alpha}} 
\ln {\cal F}^{(1)}_{\nu}(i\zeta_{\nu})
= i\eta_{\nu}\frac{m_{\nu}}{m_{{\beta}{\gamma}}}\,
\frac{1}{r_{\nu}}
\frac{ {\hat{\rm {\bf k}}}_{\nu} -{\hat {\rm {\bf r}}}_{\nu}}
{1 - {\hat{\rm {\bf k}}}_{\nu}\cdot{\hat {\rm {\bf r}}}_{\nu}}.  
\label{as1} 
\end{equation}
Note that in $\Omega_{\alpha}$ radius $\;r_{\alpha}$ is limited {\it a 
priori}
(more strictly, it is allowed to grow but slower than $\rho_{\alpha}$). 
That is why we cannot use an
asymptotic expansion in terms of  $1/\zeta_{\alpha}$ in the asymptotic 
region 
$\Omega_{\alpha}$. 
Eqs (\ref{tfnu1}),\, (\ref{sss1}), (\ref{ss1}) and (\ref{ss2}) are 
valid only in $\Omega_{\alpha}$, 
while Eqs (\ref{tldt1}), (\ref{as2})  and (\ref{as1})
are valid both in $\Omega_{0}$ and $\Omega_{\alpha}$. 

Thus we reduced a three-body problem in the asymptotic domain $\Omega_{\alpha}$
to a two-body problem: we need to find a solution of Eq. (\ref{psi3}), 
which describes the relative motion of particles $\beta$ and $\gamma$ in 
the presence of the third particle $\alpha$, which is far away, but it 
still distorts the relative motion of particles $\beta$ and $\gamma$ due 
to the long-range Coulomb interaction. This distortion results in 
the dependence of 
$\varphi_{\alpha}^{(11)}({\rm{\bf{r}}}_{\alpha},\,{\brho}_{\alpha})$ 
on $\brho_{\alpha}$. When $\rho_{\alpha}$ increases 
this distortion should be weakened. Hence, 
$\varphi_{\alpha}^{(11)}({\rm{\bf{r}}}_{\alpha},\,
{\brho}_{\alpha})$ actually depends on  $1/{\rho}_{\alpha}$ and 
\be
{\mbox{\boldmath{$\nabla$}}}_{{\brho}_{\alpha}}\,
\varphi_{\alpha}^{(11)}
({\rm{\bf{r}}}_{\alpha},\,{\brho}_{\alpha}) \sim 
\frac{1}{\rho_{\alpha}^{2}}. 
\label{gradvarphi11}
\ee
Because of that we may drop the second and sixth terms in Eq. 
(\ref{psi3}) and rewrite it in the form
\begin{eqnarray}
\lbrack{\frac{1}{2\mu_{\alpha}}}{{\mbox{\boldmath{$\Delta$}}}_{{\rm{\bf{r
}}}_{\alpha}}} +
i\,{\frac{1}{\mu_{\alpha}}}    
\,{\rm{\bf{k}}}_{\alpha}{\cdot}
{\mbox{\boldmath{$\nabla$}}}_{{\rm{\bf{r}}}_{\alpha}} +
i\,{\frac{1}{M_{\alpha}}} \,
{\rm{\bf{q}}}_{\alpha}{\cdot}
{\mbox{\boldmath{$\nabla$}}}_{{\brho}_{\alpha}} +
{\frac{1}{\mu_{\alpha}}}\,{\sum_{\nu=\beta,\gamma}} 
{\mbox{\boldmath{$\nabla$}}}_{{\rm{\bf r}}_{\alpha}}
\ln {\cal F}^{(1)}_{\nu}(i\zeta_{\nu}){\cdot}
{\mbox{\boldmath{$\nabla$}}}_{{\rm{\bf r}}_{\alpha}}\nonumber \\
-V_{\alpha}({\rm{\bf{r}}}_{\alpha})
+{\frac{1}{\mu_{\alpha}}}\, 
{\mbox{\boldmath{$\nabla$}}}_{{\rm{\bf{r}}}_{\alpha}}
\ln {\cal F}^{(1)}_{\beta}(i\zeta_{\beta}){\cdot}
{\mbox{\boldmath{$\nabla$}}}_{{\rm{\bf{r}}}_{\alpha}}
\ln {\cal F}^{(1)}_{\gamma}(i\zeta_{\gamma})\nonumber \\
+{\frac{1}{M_{\alpha}}}\,
{\mbox{\boldmath{$\nabla$}}}_{{\brho}_{\alpha}}
\ln {\cal F}^{(1)}_{\beta}(i\zeta_{\beta}){\cdot}
{\mbox{\boldmath{$\nabla$}}}_{{\brho}_{\alpha}}
\ln {\cal F}^{(1)}_{\gamma}(i\zeta_{\gamma})\rbrack 
\varphi_{\alpha}^{(11)}({\rm{\bf{r}}}_{\alpha},\,
{\brho}_{\alpha})=0. \label{psi4}
\end{eqnarray}
The last two terms are of $O(1/\rho_{\alpha}^{2})$.
Note that to satisfy this equation up to terms of 
$O(1/\rho_{\alpha}^{3})$ all the terms 
of $O(1/\rho_{\alpha}^{2})$ must be compensated.
Taking into account Eqs (\ref{tfnu1}) and (\ref{ss1}) we can rewrite  
Eq. (\ref{psi4}) as
\begin{eqnarray}
&&\lbrack 
\frac{1}{2\mu_{\alpha}}\,{\mbox{\boldmath{$\Delta$}}}_{{\rm{\bf 
r}}_{\alpha}}+
i\,\frac{1}{\mu_{\alpha}}\,{\rm{\bf k}}^{(11)}_{\alpha}
({\brho}_{\alpha}){\cdot}
{\mbox{\boldmath{$\nabla$}}}_{{\rm{\bf r}}_{\alpha}}
+i\,\frac{1}{M_{\alpha}} \,{\rm{\bf q}}_{\alpha} \cdot
{\mbox{\boldmath{$\nabla$}}}_{{\brho}_{\alpha}} \nonumber\\
&&+ \frac{1}{\mu_{\alpha}}\,{\sum_{\nu=\beta,\gamma}}
\frac{m_{\nu}^{2}}{m_{\beta \,\gamma}^{2}} 
({\rm{\bf r}}_{\alpha} \cdot 
{\mbox{\boldmath{$\nabla$}}}_{{\brho}_{\alpha}})\,
({\mbox{\boldmath{$\nabla$}}}_{{\brho}_{\alpha}}\,  
\ln {\tilde {\cal F}}^{(1)}_{\nu}(i\zeta_{\nu\,\alpha})
{\cdot} {\mbox{\boldmath{$\nabla$}}}_{{\rm{\bf r}}_{\alpha}})
 - V_{\alpha}({\rm{\bf r}}_{\alpha})\nonumber\\ 
&&+ (\epsilon_{\beta\,\alpha}\,\epsilon_{\gamma\,\alpha}\,
\frac{1}{m_{\beta\,\gamma}}+\frac{1}{M_{\alpha}})\,
{\mbox{\boldmath{$\nabla$}}}_{{\brho}_{\alpha}}\,
\ln{\cal F}^{(1)}_{\beta}(i\zeta_{\beta\,\alpha}){\cdot}
{\mbox{\boldmath{$\nabla$}}}_{{\brho}_{\alpha}}\,
\ln{\cal F}^{(1)}_{\gamma}(i\zeta_{\gamma\,\alpha})\rbrack \nonumber\\
&&\times \varphi_{\alpha}^{(11)}({\rm{\bf r}}_{\alpha},\,
{\brho}_{\alpha})=O(1/\rho_{\alpha}^{3}).
\label{psi5}
\end{eqnarray}
We introduced here a new local momentum 
\begin{equation}
{{\rm{\bf{k}}}}^{(11)}_{\alpha} = {\rm{\bf{k}}}_{\alpha}
-i {\sum_{\nu=\beta,\gamma}}\,\frac{m_{\nu}}{m_{{\beta}{\gamma}}}\,
\lbrack\, 
\epsilon_{\nu\,\alpha}\,{\mbox{\boldmath{$\nabla$}}}_{{\brho}_{\alpha}}  
\ln {\tilde {\cal F}}^{(1)}_{\nu}(i\zeta_{\nu\,\alpha}) +
i\,\eta_{\nu}^{2}\,
\frac{1}{k_{\nu}\,\rho_{\alpha}^{2}}\,
\frac{ {\hat{\rm {\bf k}}}_{\nu} - \epsilon_{\alpha \nu}
{\hat {\brho}}_{\alpha}}
{(1 -\epsilon_{\alpha \nu}{\hat{\rm {\bf k}}}_{\nu}\cdot
{\hat {\brho}}_{\alpha})^{2}} \rbrack. 
\label{tk11}
\end{equation}
Note that variables  ${\mbox{\boldmath{$\nabla$}}}_{\rm{\bf 
r}_{\alpha}}$ and 
${\mbox{\boldmath{$\nabla$}}}_{{\brho}_{\alpha}} $ are mixed up 
only in the fourth term of Eq. (\ref{psi5}).
We are looking for a solution in the form
\begin{equation}
{{\varphi}_{\alpha}^{(11)}}(\rm{\mathbf{r}}_{\alpha 
},\,{\brho}_{\alpha })
={{\varphi}_{\alpha \,(0)}^{(11)}}(%
\rm{\bf{r}}_{\alpha },\,{\brho}_{\alpha })\,(1+\frac{%
\chi (\widehat{\brho}_{\alpha })}{\rho _{\alpha }})+\frac{{%
{\varphi}_{\alpha \,(1)}^{(11)}}(\rm{\bf{r}}_{\alpha },\,%
{\brho}_{\alpha })}{\rho _{\alpha }^{2}},
\label{sol1}
\end{equation}
where $\varphi_{\alpha\,(0)}^{(11)}({\rm{\bf{r}}}_{\alpha},\,
{\brho}_{\alpha})$ is a solution of
\begin{equation}
\lbrack{\frac{1}{2\mu_{\alpha}}}{{\mbox{\boldmath{$\Delta$}}}_{{\rm{\bf{r
}}}_{\alpha}}} +
i\,{\frac{1}{\mu_{\alpha}}}\,{{\rm{\bf{k}}}}^{(11)}_{\alpha}
({\brho}_{\alpha}){\cdot}
{\mbox{\boldmath{$\nabla$}}}_{{\rm{\bf{r}}}_{\alpha}} 
- V_{\alpha}({\rm{\bf{r}}}_{\alpha})\rbrack 
\varphi_{\alpha\,(0)}^{(11)}({\rm{\bf{r}}}_{\alpha},\,
{\brho}_{\alpha})=0. \label{psi6}
\end{equation}
$\chi(\widehat{{\brho}}_{\alpha}) \sim O(1)$ and is a solution of 
the first order differential equation
\begin{eqnarray}
&&i\,{\frac{1}{M_{\alpha }}}\,\mathrm{\mathbf{q}}_{\alpha }{\cdot }{%
\mbox{\boldmath{$\nabla$}}}_{{\brho}_{\alpha }}\frac{\chi
(\widehat{{\brho}}_{\alpha })}{\rho _{\alpha }} \nonumber\\
&&=-(\epsilon _{\beta \,\alpha }\,\epsilon _{\gamma \,\alpha 
}\,\frac{1}{m_{\beta \,\gamma }}+\frac{1}{M_{\alpha }})\,
{\mbox{\boldmath{$\nabla$}}}_{{%
\brho}_{\alpha }}\ln {\tilde{\mathcal{F}}}_{\beta
}^{(1)}(i\zeta _{\beta \,\alpha }){\cdot 
}{\mbox{\boldmath{$\nabla$}}}_{{%
\brho}_{\alpha }}\ln {\tilde{\mathcal{F}}}_{\gamma
}^{(1)}(i\zeta _{\gamma \,\alpha }).
\label{chi1}
\end{eqnarray}
Finally $\varphi_{\alpha\,(1)}^{(11)}
({\rm{\bf{r}}}_{\alpha},\,{\brho}_{\alpha}) \sim O(1)$ is a solution of 
the inhomogemeous equation
\begin{eqnarray}
&&\lbrack{\frac{1}{2\mu_{\alpha}}}{{\mbox{\boldmath{$\Delta$}}}_{{\rm{\bf
{r
}}}_{\alpha}}} +
i\,{\frac{1}{\mu_{\alpha}}}\,{{\rm{\bf k}}}^{(11)}_{\alpha}{\cdot}
{\mbox{\boldmath{$\nabla$}}}_{{\rm{\bf{r}}}_{\alpha}} 
- V_{\alpha}({\rm{\bf{r}}}_{\alpha})\rbrack \,
\varphi_{\alpha\,(1)}^{(11)}({\rm{\bf{r}}}_{\alpha},\,
{\brho}_{\alpha})
=-i\,\frac{\rho_{\alpha}^{2}}{M_{\alpha}}\,{\rm{\bf q}}_{\alpha}\cdot
{\mbox{\boldmath{$\nabla$}}}_{{\brho}_{\alpha}}\,
\varphi_{\alpha\,(0)}^{(11)}({\rm{\bf r}}_{\alpha})
\nonumber \\
&&- 
{\frac{\rho_{\alpha}^{2}}{\mu_{\alpha}}}\,{\sum_{\nu=\beta,\gamma}}
\frac{m_{\nu}^{2}}{m_{\beta \,\gamma}^{2}} 
({\rm{\bf{r}}_{\alpha} \cdot 
{\mbox{\boldmath{$\nabla$}}}_{{{\brho}_{\alpha}}})\,
\mbox{\boldmath{$\nabla$}}}_{{\brho}_{\alpha}}  
\ln {\tilde {\cal F}}^{(1)}_{\nu}(i\zeta_{\nu\,\alpha})
{\cdot}
{\mbox{\boldmath{$\nabla$}}}_{{\rm{\bf{r}}}_{\alpha}}\,
\varphi_{\alpha\,(0)}^{(11)}({\rm{\bf{r}}}_{\alpha},\,
{\brho}_{\alpha}). \label{vrp1}
\end{eqnarray}
Note that all the equations (\ref{psi6}), (\ref{chi1}) and (\ref{vrp1})
are "two-body" differential equations. On the left hand side they 
contain gradients and Laplacians over only one of the variables, 
${\rm{\bf{r}}}_{\alpha}$ or ${\brho}_{\alpha}$. Therefore these equations can easily be 
solved numerically. 

Now we consider the second term of Eq. (\ref{psi2}). It satisfies the 
equation
\begin{eqnarray}
&&{\cal F}^{(2)}_{\beta}(i\zeta_{\beta})
{\cal F}^{(1)}_{\gamma}(i\zeta_{\gamma})
\lbrack{\frac{1}{2\mu_{\alpha}}}{{\mbox{\boldmath{$\Delta$}}}_{{\rm{\bf{r
}}}_{\alpha}}}+{\frac{1}{2 
M_{\alpha}}}{{\mbox{\boldmath{$\Delta$}}}_{{\brho}_{\alpha}}} + 
i\,{\frac{1}{\mu_{\alpha}}}    
\,{\rm{\bf{k}}}_{\alpha}{\cdot}
{\mbox{\boldmath{$\nabla$}}}_{{\rm{\bf{r}}}_{\alpha}} +
i\,{\frac{1}{M_{\alpha}}} \,
{\rm{\bf{q}}}_{\alpha}{\cdot}
{\mbox{\boldmath{$\nabla$}}}_{{\brho}_{\alpha}} +
\nonumber \\
&&{\frac{1}{\mu_{\alpha}}}\,\lbrack 
{\mbox{\boldmath{$\nabla$}}}_{{\rm{\bf{r}}}_{\alpha}}
\ln {\cal F}^{(2)}_{\beta}(i\zeta_{\beta}) + 
{\mbox{\boldmath{$\nabla$}}}_{{\rm{\bf{r}}}_{\alpha}}
\ln {\cal F}^{(1)}_{\gamma}(i\zeta_{\gamma})\rbrack {\cdot}
{\mbox{\boldmath{$\nabla$}}}_{{\rm{\bf{r}}}_{\alpha}} \nonumber\\
&&+ {\frac{1}{M_{\alpha}}}\,\lbrack 
{\mbox{\boldmath{$\nabla$}}}_{{\brho}_{\alpha}}
\ln {\cal F}^{(2)}_{\beta}(i\zeta_{\beta})  +  
{\mbox{\boldmath{$\nabla$}}}_{{\brho}_{\alpha}}
\ln {\cal F}^{(1)}_{\gamma}(i\zeta_{\gamma})\rbrack
{\cdot}
{\mbox{\boldmath{$\nabla$}}}_{{\brho}_{\alpha}} 
- V_{\alpha}({\rm{\bf{r}}}_{\alpha}) + \nonumber \\
&&{\frac{1}{\mu_{\alpha}}}\, 
{\mbox{\boldmath{$\nabla$}}}_{{\rm{\bf{r}}}_{\alpha}}
\ln {\cal F}^{(2)}_{\beta}(i\zeta_{\beta}){\cdot}
{\mbox{\boldmath{$\nabla$}}}_{{\rm{\bf{r}}}_{\alpha}}
\ln {\cal F}^{(1)}_{\gamma}(i\zeta_{\gamma}) + 
{\frac{1}{M_{\alpha}}}\,
{\mbox{\boldmath{$\nabla$}}}_{{\brho}_{\alpha}}
\ln {\cal F}^{(2)}_{\beta}(i\zeta_{\beta}){\cdot}
{\mbox{\boldmath{$\nabla$}}}_{{\brho}_{\alpha}}
\ln {\cal F}^{(1)}_{\gamma}(i\zeta_{\gamma})\rbrack \nonumber \\
&&\times \varphi_{\alpha}^{(21)}({\rm{\bf{r}}}_{\alpha},\,
{\brho}_{\alpha})=O(1/\rho_{\alpha}^{3}). \label{psi21}
\end{eqnarray}

Here, in the nonsingular directions (${\hat {\rm {\bf k}}}_{\nu}
\cdot {\hat {\rm {\bf r}}}_{\nu} \not=1,\,\nu \not= \alpha$)
\begin{equation}
{\cal F}_{\nu}^{(2)}(i\zeta_{\nu})\stackrel{\zeta_{\nu} \to \infty}{=}
\eta_{\nu}\,\frac{\Gamma(1 + i\,\eta_{\nu})}{\Gamma(1 - 
i\,\eta_{\nu})}\,\frac{e^{-i\,\eta_{\nu}\,\ln\,\zeta_{\nu}}}{\zeta_{\nu}}
\,
e^{i\,\zeta_{\nu}}\,
\lbrack 1 + O(\frac{1}{\zeta_{\nu}})\rbrack. \label{fas2}
\end{equation}
Also, in the nonsingular directions  for $\nu \not= \alpha$
\begin{eqnarray}
{\mbox{\boldmath{$\nabla$}}}_{{\rm{\bf{r}}}_{\alpha}}\,\ln\,{\cal 
F}_{\nu}^{(2)}(i\,\zeta_{\nu})= 
i\,{\mbox{\boldmath{$\nabla$}}}_{{\rm{\bf{r}}}_{\alpha}}\,\zeta_{\nu}
+ O(1/r_{\nu}) = i\,\frac{m_{\nu}}{m_{\beta\,\gamma}}\,k_{\nu}({\hat 
{\rm {\bf k}}}_{\nu} - {\hat {\rm {\bf r}}}_{\nu}) + O(1/r_{\nu}) 
\label{f21} \\
= i\,\frac{m_{\nu}}{m_{\beta\,\gamma}}\,k_{\nu}\,({\hat {\rm {\bf 
k}}}_{\nu} - \epsilon_{\alpha\,\nu}\,{\hat {\brho}}_{\alpha}) + 
O(1/\rho_{\alpha})   \label{cfn2}
\end{eqnarray}
and
\begin{eqnarray}
{\mbox{\boldmath{$\nabla$}}}_{{\brho}_{\alpha}} 
\ln {\cal F}^{(2)}_{\nu}(i\zeta_{\nu})
= i{\mbox{\boldmath{$\nabla$}}}_{{\brho}_{\alpha}}
\zeta_{\nu}  + O(1/r_{\nu})      
=i\epsilon_{\nu \alpha}(-k_{\nu}{\hat {\rm {\bf r}}}_{\nu} +
{\rm {\bf k}}_{\nu}) + O(1/r_{\nu})  \label{as4} \\
= i\,k_{\nu}\,({\hat {{\brho}_{\alpha}}} -
\epsilon_{\alpha\,\nu }{\hat{\rm {\bf k}}}_{\nu})  +  
O(1/\rho_{\alpha}). 
\label{asym1}
\end{eqnarray}
When deriving (\ref{psi21}) we took into account that 
\begin{equation}
({\frac{1}{2 
\mu_{\nu}}}{\mbox{\boldmath{$\Delta$}}}_{{\rm{\bf{r}}}_{\nu}} +
i\,{\frac{1}{\mu_{\nu}}}    
\,{\rm{\bf{k}}}_{\nu}{\cdot}
{\mbox{\boldmath{$\nabla$}}}_{{\rm{\bf{r}}}_{\nu}} -
V_{\nu}^{C})\,{\cal 
F}_{\nu}^{(2)}(i\zeta_{\nu})=0. \label{hyp2}
\end{equation}
To get an asymptotic equation from Eq. (\ref{psi21}) which is valid up 
to 
$O(1/\rho_{\alpha}^{3})$, all the coefficients of $O(1)$, 
$O(1/\rho_{\alpha})$ 
and $O(1/\rho_{\alpha}^{2})$ should be kept in the left-hand-side of the 
equation. 
Since in the nonsingular directions in $\Omega_{\alpha}$ region, 
${\cal F}^{(2)}_{\beta}(i\zeta_{\beta})\sim O(1/\rho_{\alpha})$ only 
coefficients of 
$O(1)$ and $O(1/\rho_{\alpha})$ in the brackets of Eq. (\ref{psi21}) 
should be left.
Taking into account Eqs (\ref{tfnu1}), (\ref{cfn2}) and (\ref{asym1}) we 
get
\begin{eqnarray}
&&\lbrack 
\frac{1}{2\mu_{\alpha}}\,{{\mbox{\boldmath{$\Delta$}}}_{{\rm{\bf 
r}}_{\alpha}}} +
i\,\frac{1}{\mu_{\alpha}}\,{{\rm{\bf k}}}^{(21)}_{\alpha}
({\brho}_{\alpha}){\cdot}
{\mbox{\boldmath{$\nabla$}}}_{{\rm{\bf r}}_{\alpha}} - 
V_{\alpha}({\rm{\bf{r}}}_{\alpha}) \nonumber\\
&&+ i\,\frac{1}{\mu_{\alpha}}\,
\frac{m_{\beta}^{2}}{m_{\beta\,\gamma}^{2}}\,k_{\beta}\,
\frac{1}{\rho_{\alpha}}\,
({\rm{\bf r}}_{\alpha} -  {\hat{\brho}}_{\alpha}\,
({\hat{\brho}}_{\alpha}{\cdot}{\rm{\bf r}}_{\alpha}))
{\cdot}{\mbox{\boldmath{$\nabla$}}}_{{\rm{\bf r}}_{\alpha}} + 
i\,\frac{1}{M_{\alpha}} \,
{{\rm{\bf q}}}^{(21)}_{\alpha}{\cdot}
{\mbox{\boldmath{$\nabla$}}}_{{\brho}_{\alpha}}  \\
&&- i\,\epsilon_{\alpha\,\beta 
}\,\frac{1}{m_{\alpha}}\,k_{\beta}\,({\hat {\rm {\bf k}}}_{\beta} - 
\epsilon_{\alpha\,\beta}\,\,{\hat {\brho}}_{\alpha})\cdot
{\mbox{\boldmath{$\nabla$}}}_{{\brho}_{\alpha}}  
\ln {\tilde {\cal F}}^{(1)}_{\gamma}(i\zeta_{\gamma\,\alpha}) 
\rbrack 
\,\varphi_{\alpha}^{(21)}({\rm{\bf{r}}}_{\alpha},\,
{\brho}_{\alpha})=O(1/\rho_{\alpha}^{2}). \nonumber
\label{psi23}
\end{eqnarray}
Here $\,{\mbox{\boldmath{$\nabla$}}}_{{\brho}_{\alpha}}
\ln{\tilde {\cal F}}^{(1)}_{\gamma}(i\zeta_{\gamma\,\alpha})\,$ is given 
by Eq. (\ref{sss1}).
We also introduced new local momenta
\begin{equation}
{{\rm{\bf k}}}^{(21)}_{\alpha}
({\brho}_{\alpha}) = {\rm{\bf k}}_{\alpha} + 
\frac{m_{\beta}}{m_{\beta\,\gamma}}\,k_{\beta}\,({\hat {\rm {\bf 
k}}}_{\beta} - \epsilon_{\alpha\,\beta}\,\,{\hat {\brho}}_{\alpha}) +
i\,(i\,\eta_{\beta} + 
1)\,\frac{m_{\beta}}{m_{\beta\,\gamma}}\,\frac{1}{\rho_{\alpha}}\,
\frac{ {\hat{\rm {\bf k}}}_{\beta} - \epsilon_{\alpha\,\beta}
{\hat {\brho}}_{\alpha}}
{1 -\epsilon_{\alpha \beta}{\hat{\rm {\bf k}}}_{\beta}\cdot
{\hat {\brho}}_{\alpha}},  \label{tldk21}
\end{equation}
and
\begin{equation}
{{\rm{\bf q}}}^{(21)}_{\alpha}
({\brho}_{\alpha})= {\rm{\bf q}}_{\alpha} + 
k_{\beta}\,({\hat {\brho}}_{\alpha} - 
\epsilon_{\alpha\,\beta}\,{\hat{\rm {\bf k}}}_{\beta}). \label{tq21}
\end{equation}
We also took into account that for 
$\nu \not=\sigma \not= 
\tau,\;\nu\not=\tau$,\;$\epsilon_{\nu\,\tau}$,\; 
$\epsilon_{\nu\,\sigma}=-1$, and
\begin{equation}
-\epsilon_{\alpha\,\gamma}\,
\frac{1}{m_{\beta\, \gamma}}\,
({\hat {\rm {\bf k}}}_{\beta} - \epsilon_{\alpha\,\beta}\,\,{\hat 
{\brho}}_{\alpha})
+ {\frac{1}{M_{\alpha}}}\,({\hat {{\brho}_{\alpha}}} 
-\epsilon_{\alpha\,\beta }{\hat{\rm {\bf k}}}_{\beta})=
- \epsilon_{\alpha\,\beta }\,\frac{1}{m_{\alpha}}\,({\hat {\rm {\bf 
k}}}_{\beta} - \epsilon_{\alpha\,\beta}\,\,{\hat {\brho}}_{\alpha}). 
\label{coeff1}
\end{equation}
We are looking for a solution of Eq. (\ref{psi23}) in the form
\begin{equation}
{\varphi}_{\alpha }^{(21)}(\rm{\bf{r}}_{\alpha },{\brho } 
_{\alpha })={\varphi}_{\alpha\,(0)}^{(21)}(\rm{\bf{r}}_{\alpha },
{\brho }_{\alpha })+\frac{{\varphi}_{\alpha 
\,(1)}^{(21)}(\rm{\bf{r}}_{\alpha },\,
{\brho}_{\alpha })}{\rho _{\alpha }},  \label{solut1}
\end{equation}
where $\varphi_{\alpha\,(0)}^{(21)}({\rm{\bf{r}}}_{\alpha},\,
{\brho}_{\alpha})$ satisfies 
\begin{equation}
\lbrack{\frac{1}{2\mu_{\alpha}}}{{\mbox{\boldmath{$\Delta$}}}_{{\rm{\bf{r
}}}_{\alpha}}} +
i\,{\frac{1}{\mu_{\alpha}}}\,{{\rm{\bf{k}}}}^{(21)}_{\alpha}
({\brho}_{\alpha}){\cdot}
{\mbox{\boldmath{$\nabla$}}}_{{\rm{\bf{r}}}_{\alpha}} 
- V_{\alpha}({\rm{\bf{r}}}_{\alpha})\rbrack 
\,\varphi_{\alpha\,(0)}^{(21)}({\rm{\bf{r}}}_{\alpha},\,
{\brho}_{\alpha})=0. \label{psi210}
\end{equation}
Finally $\varphi_{\alpha\,(1)}^{(21)}
({\rm{\bf{r}}}_{\alpha},\,{\brho}_{\alpha}) \sim O(1)$ is a solution of 
equation
\begin{eqnarray}
&&[\frac{1}{2\mu _{\alpha }}\,{{\mbox{\boldmath{$\Delta$}}} 
_{\mathrm{\mathbf{r}}_{\alpha }}}+i\,%
\frac{1}{\mu _{\alpha }}\,{{\mathrm{\mathbf{k}}}}_{\alpha}^{(21)}
({\brho }_{\alpha }){\cdot \nabla }_{\mathrm{\mathbf{r}}_{\alpha 
}}-V_{\alpha }(%
\mathrm{\mathbf{r}}_{\alpha })]{\varphi}_{\alpha \,(1)}^{(21)}(%
\mathrm{\mathbf{r}}_{\alpha },\,{\brho }_{\alpha })  \nonumber \\
&&=-[i\,\frac{1}{\mu _{\alpha }}\,\frac{%
m_{\beta }^{2}}{m_{\beta \,\gamma }^{2}}\,k_{\beta }\,
(\mathrm{\mathbf{r}}_{\alpha }-{\hat{\brho}}_{\alpha }\,({\hat{\brho}}%
_{\alpha }{\cdot }\mathrm{\mathbf{r}}_{\alpha })){\cdot \nabla 
}_{\mathrm{%
\mathbf{r}}_{\alpha }}]{\varphi}_{\alpha \,(0)}^{(21)}(\mathrm{%
\mathbf{r}}_{\alpha },\,{\brho }_{\alpha })  \nonumber \\
&&-i\,\frac{{\rho }_{\alpha }}{M_{\alpha }}{{\mathrm{%
\mathbf{q}}}}_{\alpha }^{(21)}{\cdot \nabla }_{{\rho }_{\alpha 
}}{\varphi}_{\alpha \,(0)}^{(21)}(\mathrm{\mathbf{r}}_{\alpha },\,{\brho 
}_{\alpha })   \nonumber\\
&&+i\,\epsilon_{\alpha\,\beta }\,\frac{{\rho }_{\alpha 
}}{m_{\alpha}}\,k_{\beta}\,({\hat {\rm {\bf k}}}_{\beta} - 
\epsilon_{\alpha\,\beta}\,\,{\hat {\brho}}_{\alpha})\cdot
{\mbox{\boldmath{$\nabla$}}}_{{\brho}_{\alpha}}  
\ln {\tilde {\cal F}}^{(1)}_{\gamma}(i\zeta_{\gamma\,\alpha})
{\varphi}_{\alpha \,(0)}^{(21)}(\mathrm{\mathbf{r}}_{\alpha 
},\,{\brho }_{\alpha }).    \label{vrp21}
\end{eqnarray}
Since in Eq. (\ref{vrp21}) we keep only terms of order 
$O(1/\rho_{\alpha})\;$ local momentum 
${{\rm{\bf k}}}^{(21)}_{\alpha}({\brho}_{\alpha})$ can be replaced by
\begin{equation}
{{\rm{\bf k}}}^{(21)}_{\alpha (0)}
({\brho}_{\alpha}) = {\rm{\bf k}}_{\alpha} + 
\frac{m_{\beta}}{m_{\beta\,\gamma}}\,k_{\beta}\,({\hat {\rm {\bf 
k}}}_{\beta} - \epsilon_{\alpha\,\beta}\,\,{\hat {\brho}}_{\alpha}). 
\label{tk210}
\end{equation}
A formal solution of Eq. (\ref{vrp21}) is
\begin{eqnarray}
&&\varphi_{\alpha\,(1)}^{(21)}({\rm{\bf{r}}}_{\alpha},\,
{\brho}_{\alpha})= 
\varphi_{\alpha\,(0)}^{(21)}({\rm{\bf{r}}}_{\alpha},\,
{\brho}_{\alpha})
+ e^{-{{\rm{\bf k}}}^{(21)}_{\alpha}({\brho}_{\alpha})\cdot {\rm {\bf 
r}}_{\alpha}}\,
\int\,{\rm d}\,{\rm {\bf r}}_{\alpha}'\,G({\rm {\bf r}}_{\alpha},\,{\rm 
{\bf r}}_{\alpha}')\,e^{{{\rm{\bf k}}}^{(21)}_{\alpha}({\brho}_{\alpha})
{\cdot} {\rm {\bf r}}_{\alpha}^{'}}\, \nonumber\\
&&\lbrack -[i\,\frac{1}{\mu _{\alpha }}\,\frac{%
m_{\beta }^{2}}{m_{\beta \,\gamma }^{2}}\,k_{\beta }\,
(\mathrm{\mathbf{r}}_{\alpha }^{'}-{\hat{\brho}}_{\alpha 
}\,({\hat{\brho}}%
_{\alpha }{\cdot }\mathrm{\mathbf{r}}_{\alpha }^{'})){\cdot \nabla 
}_{\mathrm{%
\mathbf{r}}_{\alpha }}]{\varphi}_{\alpha \,(0)}^{(21)}(\mathrm{%
\mathbf{r}}_{\alpha }^{'},\,{\rho }_{\alpha })\nonumber\\
&&-i\,\frac{1}{M_{\alpha}}\,\,{{\rm{\bf 
q}}}^{(21)}_{\alpha}({\brho}_{\alpha})
\cdot{\mbox{\boldmath{$\nabla$}}}_{
{\brho}_{\alpha}}\,\varphi_{\alpha\,(0)}^{(11)}({\rm{\bf 
r}}_{\alpha}^{'}) \nonumber\\
&&- i\,\epsilon_{\alpha\,\beta 
}\,\frac{1}{m_{\alpha}}\,k_{\beta}\,({\hat {\rm {\bf k}}}_{\beta} - 
\epsilon_{\alpha\,\beta}\,\,{\hat {\brho}}_{\alpha})\cdot
{\mbox{\boldmath{$\nabla$}}}_{{\brho}_{\alpha}}  
\ln {\tilde {\cal F}}^{(1)}_{\gamma}(i\zeta_{\gamma\,\alpha}^{'})\,
\varphi_{\alpha\,(0)}^{(21)}({\rm{\bf r}}_{\alpha}^{'},\,
{\brho}_{\alpha})\rbrack, \label{frmsl21}
\end{eqnarray}
Here $\varphi_{\alpha\,(0)}^{(21)}({\rm{\bf{r}}}_{\alpha},\,
{\brho}_{\alpha})$ is a solution of the homogeneous Eq. 
(\ref{psi210}).

The third equation for
$\varphi_{\alpha}^{(12)}({\rm{\bf{r}}}_{\alpha},\,
{\brho}_{\alpha})$ is obtained by substituting the third term in 
(\ref{psi2}) to (\ref{ThSE}). Following the same steps, which we 
used to derive the second  equation, or just interchanging 
$\beta \leftrightarrow \gamma $ in (\ref{psi21}) we find  
$\varphi_{\alpha}^{(12)}({\rm{\bf{r}}}_{\alpha},\,
{\brho}_{\alpha})$ in the following form:
\begin{equation}
{\varphi}_{\alpha }^{(12)}(\rm{\bf{r}}_{\alpha },{\brho } 
_{\alpha})
={\varphi}_{\alpha \,(0)}^{(12)}(\rm{\bf{r}}_{\alpha }, {\brho 
}_{\alpha })
+\frac{{\varphi}_{\alpha \,(1)}^{(12)}
(\rm{\bf{r}}_{\alpha },\,{\brho}_{\alpha })}{\rho _{\alpha }},
\label{sol12}
\end{equation}
where $\varphi_{\alpha\,(0)}^{(12)}({\rm{\bf{r}}}_{\alpha},\,
{\brho}_{\alpha})$ is a solution of
\begin{equation}
\lbrack{\frac{1}{2\mu_{\alpha}}}{{\mbox{\boldmath{$\Delta$}}}_{{\rm{\bf{r
}}}_{\alpha}}} +
i\,{\frac{1}{\mu_{\alpha}}}\,{{\rm{\bf{k}}}}^{(12)}_{\alpha}
({\brho}_{\alpha}){\cdot}
{\mbox{\boldmath{$\nabla$}}}_{{\rm{\bf{r}}}_{\alpha}} 
- V_{\alpha}({\rm{\bf{r}}}_{\alpha})\rbrack 
\, \varphi_{\alpha\,(0)}^{(12)}({\rm{\bf{r}}}_{\alpha},\,
{\brho}_{\alpha})=0. \label{psi120}
\end{equation}
We can derive a similar to Eq. (\ref{vrp21}) equation  for
${\varphi}_{\alpha \,(1)}^{(12)}(\rm{\bf{r}}_{\alpha },\,{\rho}_{\alpha 
})$ 
which has a formal solution
\begin{eqnarray}
&&\varphi_{\alpha\,(1)}^{(12)}({\rm{\bf{r}}}_{\alpha},\,
{\brho}_{\alpha})= 
\varphi_{\alpha\,(0)}^{(12)}({\rm{\bf{r}}}_{\alpha},\,
{\brho}_{\alpha}) +
e^{-{{\rm{\bf k}}}^{(21)}_{\alpha}({\brho}_{\alpha})\cdot {\rm {\bf 
r}}_{\alpha}}\,
\int\,{\rm d}\,{\rm {\bf r}}_{\alpha}'\,G({\rm {\bf r}}_{\alpha},\,{\rm 
{\bf r}}_{\alpha}')\,e^{{ {\rm{\bf k}}}^{(12)}_{\alpha}
({\brho}_{\alpha}){\cdot} {\rm {\bf r}}_{\alpha}^{'}}\, \nonumber\\
&&\lbrack -[i\,\frac{1}{\mu _{\alpha }}\,\frac{%
m_{\beta }^{2}}{m_{\beta \,\gamma }^{2}}\,k_{\beta }\,
(\mathrm{\mathbf{r}}_{\alpha }^{'}-{\hat{\brho}}_{\alpha 
}\,({\hat{\brho}}%
_{\alpha }{\cdot }\mathrm{\mathbf{r}}_{\alpha }^{'})){\cdot \nabla 
}_{\mathrm{%
\mathbf{r}}_{\alpha }}]{\varphi}_{\alpha \,(0)}^{(12)}(\mathrm{%
\mathbf{r}}_{\alpha }^{'},\,{\rho }_{\alpha })\nonumber\\
&&-i\,\frac{1}{M_{\alpha}}\,\,{{\rm{\bf q}}}^{(12)}_{\alpha}
({\brho}_{\alpha})\cdot{\mbox{\boldmath{$\nabla$}}}_{
{\brho}_{\alpha}}\,\varphi_{\alpha\,(0)}^{(11)}({\rm{\bf 
r}}_{\alpha}^{'}) \nonumber\\
&&- i\,\epsilon_{\alpha\,\beta 
}\,\frac{1}{m_{\alpha}}\,k_{\beta}\,({\hat {\rm {\bf k}}}_{\beta} - 
\epsilon_{\alpha\,\beta}\,\,{\hat {\brho}}_{\alpha})\cdot
{\mbox{\boldmath{$\nabla$}}}_{{\brho}_{\alpha}}  
\ln {\tilde {\cal 
F}}^{(1)}_{\gamma}(i\zeta_{\gamma\,\alpha}^{'})\,
\varphi_{\alpha\,(0)}^{(12)}({\rm{\bf 
r}}_{\alpha}^{'},\,{\brho}_{\alpha})\rbrack. 
\label{frmsl12}
\end{eqnarray}
The fourth equation can be derived after substituting the last term of 
Eq. 
(\ref{psi2}) into Eq.(\ref{ThSE}) and it is automatically satisfied 
up to the terms of order $O(1/{\rho_{\alpha}}^{3})$ in $\Omega_{\alpha}$ 
because the product ${\cal F}^{(2)}_{\beta}(i\zeta_{\beta})
{\cal F}^{(1)}_{\gamma}(i\zeta_{\gamma})=O(1/{\rho_{\alpha}}^{2})$. 
The fourth term in Eq. (\ref{psi2}) leads to an equation for 
${\varphi}_{\alpha}^{(22)}({\rm{\bf{r}}}_{\alpha},\,
{\brho}_{\alpha})$:
\begin{eqnarray}
\mathcal{F}_{\beta }^{(2)}(i\zeta _{\beta })\mathcal{F}_{\gamma
}^{(2)}(i\zeta _{\gamma })[{\frac{1}{2\mu _{\alpha 
}}}{{\mbox{\boldmath{$\Delta$}}} _{\mathrm{%
\mathbf{r}}_{\alpha }}}
+{\frac{1}{2M_{\alpha }}}{{\mbox{\boldmath{$\Delta$}}} 
_{\mathrm{\mathbf{\rho }}_{\alpha }}}
+i\,{\frac{1}{\mu _{\alpha }}}\,\mathrm{\mathbf{k}}%
_{\alpha }{\cdot \mbox{\boldmath{$\nabla$}} 
}_{\mathrm{\mathbf{r}}_{\alpha }}+i\,{%
\frac{1}{M_{\alpha }}}\,\mathrm{\mathbf{q}}_{\alpha }{\cdot 
\mbox{\boldmath{$\nabla$}} 
}_{{\brho}_{\alpha }}\nonumber\\
+{\frac{1}{\mu _{\alpha }}}\,[{\mbox{\boldmath{$\nabla$}} 
}_{\mathrm{\mathbf{r}}%
_{\alpha }}\ln \mathcal{F}_{\beta }^{(2)}(i\zeta _{\beta })+{%
\mbox{\boldmath{$\nabla$}} }_{\mathrm{\mathbf{r}}_{\alpha }}\ln 
\mathcal{F}_{\gamma
}^{(2)}(i\zeta _{\gamma })]{\cdot \mbox{\boldmath{$\nabla$}} 
}_{\mathrm{\mathbf{r}}%
_{\alpha }}\nonumber\\
+{\frac{1}{M_{\alpha }}}\,[{\mbox{\boldmath{$\nabla$}} }_{{\brho}%
_{\alpha }}\ln \mathcal{F}_{\beta }^{(2)}(i\zeta _{\beta })+{%
\mbox{\boldmath{$\nabla$}} }_{{\brho}_{\alpha }}\ln \mathcal{F}%
_{\gamma }^{(2)}(i\zeta _{\gamma })]
{\cdot \mbox{\boldmath{$\nabla$}} }_{{\brho}_{\alpha }}\nonumber\\
-V_{\alpha }(\mathrm{\mathbf{r}}_{\alpha })
+{\frac{1}{\mu _{\alpha }}}\,{\mbox{\boldmath{$\nabla$}} 
}_{\mathrm{\mathbf{r}}%
_{\alpha }}\ln \mathcal{F}_{\beta }^{(2)}(i\zeta _{\beta }){\cdot
\mbox{\boldmath{$\nabla$}} }_{\mathrm{\mathbf{r}}_{\alpha }}\ln 
\mathcal{F}_{\gamma
}^{(2)}(i\zeta _{\gamma })\\
+{\frac{1}{M_{\alpha }}}\,{\mbox{\boldmath{$\nabla$}} }_{%
{\brho}_{\alpha }}\ln \mathcal{F}_{\beta }^{(2)}(i\zeta
_{\beta }){\cdot \mbox{\boldmath{$\nabla$}} }_{{\brho}_{\alpha }}\ln 
\mathcal{F}_{\gamma }^{(2)}(i\zeta _{\gamma })]\times 
{\varphi}_{\alpha }^{(22)}(\mathrm{\mathbf{r}}_{\alpha },{\brho}%
_{\alpha })=O(1/\rho _{\alpha }^{3}).
\nonumber
\label{psi22}
\end{eqnarray}
Using the same arguments we have used before, we may drop all the terms 
containing derivatives over ${\mathbf{\rho }}_{\alpha }$ when looking 
for a solution in leading order. 
Then the equation for ${\varphi}_{\alpha }^{(22)}$ reduces to 
\begin{equation}
\lbrack{\frac{1}{2\mu_{\alpha}}}{{\mbox{\boldmath{$\Delta$}}}_{{\rm{\bf{r
}}}_{\alpha}}} +
i\,{\frac{1}{\mu_{\alpha}}}\,{{\rm{\bf{k}}}}^{(22)}_{\alpha}
({\brho}_{\alpha}){\cdot}
{\mbox{\boldmath{$\nabla$}}}_{{\rm{\bf{r}}}_{\alpha}} 
- V_{\alpha}({\rm{\bf{r}}}_{\alpha})\rbrack 
\,\varphi_{\alpha}^{(22)}({\rm{\bf{r}}}_{\alpha},\,
{\brho}_{\alpha})=0, \label{psifunct22}
\end{equation}
with a local momentum
\begin{equation}
{{\mathrm{\mathbf{k}}}}_{\alpha }^{(22)}({\brho}%
_{\alpha })=\mathrm{\mathbf{k}}_{\alpha }+\sum\limits_{\nu =\beta 
,\gamma }%
\frac{m_{\nu }}{m_{\beta \,\gamma }}\,k_{\nu 
}\,({\hat{\mathrm{\mathbf{k}}}}%
_{\nu }-\epsilon _{\alpha \,\nu }\,\,{\hat{\brho}}%
_{\alpha }).  \label{tldk22}
\end{equation}
If $V_{\alpha}$ is a pure Coulomb potential, 
$V_{\alpha}=V_{\alpha}^{C}$, then Eqs (\ref{psi6}), (\ref{psi210}),
(\ref{psi120}), (\ref{psifunct22}) have the following solution
\begin{eqnarray}
{\varphi}_{\alpha}^{(ij)}(\mathrm{\mathbf{r}}_{\alpha },
\,{\brho }_{\alpha })=N_{\alpha }^{(ij)}(\,{\brho}%
_{\alpha })F(-i\eta _{\alpha }^{(ij)}(\,{\brho}_{\alpha
}),1;i\zeta ^{(ij)}(\,{\brho}_{\alpha })),
\end{eqnarray}
Here, $i=1,2$; $j=1,2$ and $N_{\alpha }^{(ij)}(\,{\brho}_{\alpha })$ 
is defined as
\be
N_{\alpha }^{(ij)}(\,{\brho}_{\alpha })=e^{-\pi 
\eta_{\alpha }^{(ij)}(\,{\brho}_{\alpha })/2}\Gamma (1+i
\eta_{\alpha }^{(ij)}(\,{\brho}_{\alpha })), 
\label{normfactor1}
\ee
where
$\eta _{\alpha }^{(ij)}(\,{\brho}_{\alpha })=\frac{z_{\beta
}z_{\gamma }\,e^{2}\mu _{\alpha }}{k_{\alpha }^{(ij)}(\,{\brho}_{\alpha 
})}$,
and 
$\zeta ^{(ij)}(\,{\brho}_{\alpha })=k_{\alpha }^{(ij)}(\,%
{\brho}_{\alpha })r_{\alpha }-{\mathbf{k}}_{\alpha
}^{(ij)}(\,{\brho}_{\alpha })\cdot {\mathbf{r}}_{\alpha }$.

If $V_{\alpha}$ is not a pure Coulomb potential, then the differential 
equations above, which parametrically depend on ${\brho}_{\alpha}$, 
should be solved numerically. 
Since all equations are of the two-body type, numerical methods are well 
developed and have been in use for a long time. 
They can be applied to solve the differential equations above as well. 
All the solutions found this way are valid in all directions of the 
asymptotic region
$\Omega_{\alpha}$ except for singular directions.

Thus, returning to Eq. (\ref{psi2}) we can claim that, having derived
all four wave functions 
${\varphi}_{\alpha \,(1)}^{(ij)}(\rm{\bf{r}}_{\alpha },\,{\rho}_{\alpha 
}),\,i,j=1,2$, we know the asymptotic behavior of the three-body 
incident wave of the scattering wave function of the first 
type in the asymptotic region $\Omega_{\alpha}$ up to terms  $O(1/{\rho}_{\alpha}^{3})$.

\section{Generalized asymptotic scattering wave function valid in all 
regions $\Omega_{\nu},\;\nu=\alpha,\, \beta,\, \gamma $}

Now we are in position to present a generalized 
asymptotic scattering wave function which satisfies the \SE up to second 
order and which is valid in all the asymptotic regions: 
\begin{eqnarray}
\Psi_{{\mathbf{k}}_{\alpha }{\mathbf{q}}_{\alpha 
}}^{(\alpha\beta\gamma)(+)}
({\mathbf{r}}_{\alpha },{\brho}_{\alpha })\equiv e^{i%
{\mathbf{k}}_{\alpha }\cdot {\mathbf{r}}_{\alpha }+i%
{\mathbf{q}}_{\alpha }\cdot {\brho}_{\alpha }}
\varphi_{\widetilde{\mathbf{k}}_{\alpha } }({\mathbf{r}}_{\alpha })
\varphi_{\widetilde{\mathbf{k}}_{\beta }}({\mathbf{r}}_{\beta })
\varphi_{\widetilde{\mathbf{k}}_{\gamma }}({\mathbf{r}}_{\gamma }).
\label{genwave}
\end{eqnarray}
After substituting (\ref{genwave}) into (\ref{ThSE}) and dropping the 
higher order terms we get,
\begin{eqnarray}
&&\{E-T_{{\mathbf{r}}_{\alpha }}-T_{\vec{\rho}_{\alpha 
}}-V\}[e^{i{\mathbf{k}%
}_{\alpha }\cdot \vec{r}_{\alpha }+i{\mathbf{q}}_{\alpha }\cdot {\brho}%
_{\alpha }}\varphi _{\widetilde{{\mathbf{k}}}_{\alpha }}({\mathbf{r}}%
_{\alpha })\varphi _{\widetilde{{\mathbf{k}}}_{\beta 
}}({\mathbf{r}}_{\beta
})\varphi _{\widetilde{{\mathbf{k}}}_{\gamma }}({\mathbf{r}}_{\gamma })] 
\notag \\
&&=e^{i{\mathbf{k}}_{\alpha }\cdot {\mathbf{r}}_{\alpha 
}+i{\mathbf{q}}%
_{\alpha }\cdot {\brho}_{\alpha }}\varphi 
_{\widetilde{{\mathbf{k}}}_{\alpha
}}({\mathbf{r}}_{\alpha })\varphi _{\widetilde{{\mathbf{k}}}_{\gamma 
}}({%
\mathbf{r}}_{\gamma }) 
[\frac{\bigtriangleup _{{\mathbf{r}}_{\alpha }}}{2\mu _{\alpha 
}}+\frac{i%
\widetilde{{\mathbf{k}}}_{\alpha }\cdot \bigtriangledown _{{\mathbf{r}}%
_{\alpha }}}{\mu _{\alpha }}-V_{\alpha }]\varphi 
_{\widetilde{{\mathbf{k}}}%
_{\alpha }}({\mathbf{r}}_{\alpha }) \\
&&+e^{i{\mathbf{k}}_{\beta }\cdot {\mathbf{r}}_{\beta 
}+i{\mathbf{q}}_{\beta
}\cdot {\brho}_{\beta }}\varphi _{\widetilde{{\mathbf{k}}}_{\alpha }}({%
\mathbf{r}}_{\alpha })\varphi _{\widetilde{{\mathbf{k}}}_{\gamma 
}}({\mathbf{%
r}}_{\gamma })  
[\frac{\bigtriangleup _{{\mathbf{r}}_{\beta }}}{2\mu _{\beta }}+\frac{i%
\widetilde{{\mathbf{k}}}_{\beta }\cdot \bigtriangledown _{{\mathbf{r}}%
_{\beta }}}{\mu _{\beta }}-V_{\beta }]\varphi 
_{\widetilde{{\mathbf{k}}}%
_{\beta }}({\mathbf{r}}_{\beta })  \notag \\
&&+e^{i{\mathbf{k}}_{\gamma }\cdot {\mathbf{r}}_{\gamma }+i{\mathbf{q}}%
_{\gamma }\cdot {\brho}_{\gamma }}\varphi 
_{\widetilde{{\mathbf{k}}}_{\alpha
}}({\mathbf{r}}_{\alpha })\varphi _{\widetilde{{\mathbf{k}}}_{\beta 
}}({%
\mathbf{r}}_{\beta }) 
\lbrack \frac{\bigtriangleup _{{\mathbf{r}}_{\gamma }}}{2\mu
_{\gamma }}+\frac{i\widetilde{{\mathbf{k}}}_{\gamma }\cdot 
\bigtriangledown
_{{\mathbf{r}}_{\gamma }}}{\mu _{\gamma }}\bigskip -V_{\gamma }]\varphi 
_{%
\widetilde{{\mathbf{k}}}_{\gamma }}({\mathbf{r}}_{\gamma })  \notag \\
&&=\left\{ 
\begin{array}{c}
O(\frac{1}{r_{\alpha }^{2}},\frac{1}{r_{\beta }^{2}},\frac{1}{r_{\gamma 
}^{2}%
}),\;{\mathbf{r}}_{\alpha },{\mathbf{r}}_{\beta },{\mathbf{r}}_{\gamma 
}\in
\Omega _{0} \\ 
O(\frac{1}{r_{\beta }^{2}},\frac{1}{r_{\gamma 
}^{2}}),\;{\mathbf{r}}_{\beta
},{\mathbf{r}}_{\gamma }\in \Omega _{\alpha } \\ 
O(\frac{1}{r_{\alpha }^{2}},\frac{1}{r_{\gamma }^{2}}),\;{\mathbf{r}}%
_{\alpha },{\mathbf{r}}_{\gamma }\in \Omega _{\beta } \\ 
O(\frac{1}{r_{\alpha }^{2}},\frac{1}{r_{\beta 
}^{2}}),\;{\mathbf{r}}_{\alpha
},{\mathbf{r}}_{\beta }\in \Omega _{\gamma }
\end{array}\right. ,  
\label{generwf1}
\end{eqnarray}
where the local momentum is given by
\begin{eqnarray}
\widetilde{\mathbf{k}}_{\nu }={\mathbf{k}}_{\nu }-i\sum\limits_{\tau
=\alpha,\beta ,\gamma }(1-\delta_{\nu,\tau})\bigtriangledown 
_{{\mathbf{r}}_{\nu }}\ln \varphi_{\widetilde{\mathbf{k}}_{\tau}}. 
\label{locmoment1}
\end{eqnarray}
In the asymptotic region $\Omega_{0}$, each local momentum, 
$\widetilde{\mathbf{k}}_{\nu }$, can be replaced by the corresponding 
asymptotic momentum, ${\mathbf{k}}_{\nu }$. 
In the asymptotic region $\Omega_{\alpha}$, 
Eq.(\ref{generwf1}) reduces to the one quasi-two-particle differential 
equations: 
\begin{eqnarray}
\lbrack \frac{\bigtriangleup _{{\mathbf{r}}_{\alpha }}}{2\mu
_{\alpha }}+\frac{i\widetilde{\mathbf{k}}_{\alpha }\cdot 
\bigtriangledown _{%
{\mathbf{r}}_{\alpha }}}{\mu _{\alpha }}-V_{\alpha}]
\varphi _{\widetilde{\mathbf{k}}_{\alpha }}({\mathbf{r}}_{\alpha })
=O(\frac{1}{r_{\beta}},\frac{1}{r_{\gamma}}).
\label{twoparteq1}
\end{eqnarray}
Solution of this equation is evident and provides the 
Coulomb-nuclear scatttering wave function with the local momentum 
$\widetilde{\mathbf{k}}_{\alpha}$.
Similarly we can get the asymptotic solution in leading order in the 
other two asymptotic regions $\Omega_{\beta}$ and $\Omega_{\gamma}$.

\section{Conclusion}

We derived the three-body asymptotic incident wave, which 
satisfies the {\SE\,} in the asymptotic region 
$\Omega_{\nu},\;\nu=\alpha,\,\beta,\,\gamma$ up to terms of order 
$1/{\rho}_{\nu}^{3}$. 
This asymptotic incident wave gives the leading asymptotic terms of the 
three-body scattering wave function of the first type and 
is an extention of the asymptotic wave function derived in 
\cite{AM93,ML96}. 
Equivalently, similar wave functions satisfy the {\SE\,} up to 
$O(1/{\rho}_{\nu}^{3}),\,\nu=\beta,\,\gamma$. 
It is worth mentioning that the asymptotic solution satisfying the {\SE,}
in the asymptotic region $\Omega_{\nu}$ up to the 
$O(1/{\rho}_{\nu}^{2})$ can be found analytically \cite{AM93, ML96}. To 
find an asymptotic solution satisfying the {\SE\,} in  
$\Omega_{\nu}$ up to terms of $O(1/{\rho}_{\nu}^{3})$ we need to solve 
two-body type differential equations numerically. The next order term in 
the asymptotic three-body scattering wave function represents the 
outgoing $3\; particles \to 3\; particles$ scattered wave and 
has been given in \cite{KMSB04}.
  
The resulting asymptotic solution provides extended boundary conditions 
in all the asymptotic regions and can be used in the direct numerical 
solution of the {\SE\,} or in approximate perturbation calculations as a leading asymptotic 
term of the three-body scattering wave function.

\begin{acknowledgments}
This work was supported by the U.\ S.\ DOE under Grant No.\
DE-FG03-93ER40773, by NSF Award No.\ PHY-0140343 and the Australian Research Council.
\end{acknowledgments}


\end{document}